# Solving Large Traveling Salesman Problems (TSPs) by a Recursive Clustering Algorithm and a Scalable FPGA-Based Implementation

Junqi Huang, Hawraa Abbas Almurieb, T.Nandha Kumar, Haider A.F. Almurib and Fabrizio Lombardi, Fellow IEEE

***Abstract*— The Traveling Salesman Problem (TSP) continues to attract significant research interest due to its critical role in various applications. This paper introduces a recursive clustering approach that divides cities into a limited number of clusters, each containing up to five cities and its own centroid. Constrained TSP and simulated annealing techniques are employed to route cities within each cluster, using the centroids of neighboring clusters as entry and exit points for the routing process. This method offers the benefit of producing accurate and cost-effective routing solutions, due to the reduced number of cities in each cluster. The connections between cluster centroids are established using simulated annealing. The FPGA-based proposed hardware implementation demonstrates the feasibility of routing a large number of cities, as the approach leverages memory to store cluster information. Consequently, the method is independent of the FPGA's logic hardware, and its scalability depends on the FPGA's memory capacity. Furthermore, distance calculations using approximate methods for the FPGA execution are compared and the squared Euclidean is chosen due to its low resources' utilization. Compared to previous methods, the proposed approach can handle more cities for routing, offering better scalability and a higher operating frequency.**

## I. INTRODUCTION

COMBINATIONAL optimization focuses on developing efficient methods to optimize cost functions for problems with numerous independent variables (N) [2,3]. Typically, such problems belong to a well-known class of nondeterministic polynomial-time complete (NP-complete) problems, which have been extensively studied [3]. A prime example of a combinatorial optimization problem (COP) is the travelling salesman problem (TSP). TSP involves planning a route that visits N cities while minimizing travel costs and returning to the origin or starting location [3]. These problems are also prevalent in scheduling and design contexts, including layout routing of VLSI chips, drug delivery, robot path planning, gaming, machine learning, and natural language processing. Although conventional computer architecture has sought optimal solutions to NP-complete problems, it struggles with the complexity of hardware and the execution time required to find such solutions, and results are possible only for small values of N because computational effort increases exponentially with N. Consequently, it fails to meet the demands of many applications [3,4].

To address this issue, two strategies have been considered. The first strategy decomposes the problem into subproblems with powers of N corresponding to computational requirements (i.e., a small number of N), while ensuring that the subproblems are naturally disjoint. These subproblems are solved, and their solutions are integrated to obtain the overall solution to the original problem [5]. Alternatively, the second strategy reconfigures a system's known configuration to improve the cost function [6,7]. This rearrangement continues until the optimal cost function is achieved. This method involves several iterations and typically results in a local optimum rather than a global optimum. Therefore, addressing the issue of local optima is required to obtain the best cost function. At this juncture, it is worth noting that finding an optimum solution to the NP-complete problem resembles the annealing method in statistical mechanics. As shown by Metropolis in [8], the annealing process is different from the iterative method while supporting the adaptive heuristic method. Furthermore, in recent years, new hardware architectures known as quantum annealers [9], (based on quantum annealing) and CMOS annealers [10] (based on simulated annealing (SA)) have emerged. These architectures map COPs to the ground-state search of an Ising model [10][14]; among them, most COPs can be effectively mapped to an Ising model and solved using heuristic methods, such as SA [15], parallel tempering [16], genetic algorithms [17], and parallel annealing [14][18-20].

A method [14] employs a parallel annealing-based Ising machine to solve the TSP. However, this approach encounters

Junqi Huang is with School of Opto-electronic and Communication Engineering, Xiamen University of Technology, Xiamen, China. Email: huangjunqi@xmut.edu.cn.

Hawraa Abbas Almurieb is with Department of Mathematics, College of Education for Pure Sciences, University of Babylon, Iraq. E-mail: pure.hawraa.abbas@uobabylon.edu.iq.

T.Nandha Kumar, Haider A.F. Almurib, are with Department of Electrical and Electronic Engineering, University of Nottingham, Malaysia. E-mail: nandhakumaar.t@nottingham.edu.my; haider.abbas@nottingham.edu.my.

Fabrizio Lombardi is with Department. of Electrical and Computer Engineering., Northeastern University, Boston, MA 02115, USA. E-mail: lombardi@ece.neu.edu.

Corresponding Author: Haider A.F. Almurib, are with Department of Electrical and Electronic Engineering, University of Nottingham, Malaysia. E-mail: haider.abbas@nottingham.edu.my.

issues with local minimum states when addressing the constrained TSP. Moreover, the authors performed a piecewise linear approximation of the sigmoidal function for lower Ising energies.

A linear-approximation method was employed in [25] as a hardware approach for an Ising machine that uses an enhanced parallel annealing algorithm. This scheme has improved the rate of accumulation and reduced hardware requirements; however, it still requires time-consuming, hardware-intensive, and complex mathematical computations.

Another method (referred to as improved PA (IPA)) uses an exponential temperature function with a dynamic offset to solve TSP [20]. This method decomposes the number of clusters. The maximum number of cities in a cluster at different layers is fixed, and as the number of layers increases, the number of cities within the cluster decreases. Centroids of the clusters are determined at each level and interconnected through the TSP algorithm; although this method employs a heuristic approach to solve TSP, it effectively addresses the TSP for a limited number of cities. Similarly, the method in [18] presents a clustering approach that uses a fixed maximum number of cities per cluster for a given layer. As the number of layers increases, the maximum number of cities in each cluster also decreases. The method does not indicate the connections between clusters; instead, it uses cluster centroids and links them via the TSP algorithm. Although the results of this method indicate that it can be utilized for significantly fewer than one hundred cities, it does not demonstrate its effectiveness beyond the city bound.

Another issue arising in TSP is related to determining the metric used to calculate the distances between cities; most methods use the Euclidean metric to determine the exact distance between two cities based on their coordinates. Other metrics may be used to approximate distances, particularly for grids and discrete networks, such as Manhattan, Chebyshev, and Octile Metrics. Moreover, the Octile metric has received considerable attention across a range of applications, particularly for shortest-path approximation problems. [23] has generated subgoal graphs that can be used to quickly find the shortest path using an eight-neighbor grid. Additionally, in the context of pathfinding problems, [22] employed the Octile metric to reduce computational costs associated with bidding in multirobot routing. Another use of the Octile metric is in General Purpose Units to handle operations between graphs. In [24], the TSP is one of the experiments that used the Octile metric to control the implementation run time. However, the square Euclidean has proven to be more amenable to hardware.

Based on the findings from the previously described techniques for addressing the TSP, the proposed recursive method uses five clusters at each level, increasing the number of levels until each cluster contains at most five cities. This strategy improves parallel processing efficiency and reduces the execution time for discovering both local and global optimal solutions. Moreover, the method uses centroids in the initial layer and links them via the TSP algorithm; in subsequent layers, it applies a constrained TSP (CTSP) (or Path TSP) technique, which substantially improves execution speed and significantly reduces overall distance in the TSP. In addition, a novel approximation technique has been developed for the probabilistic activation function. This method achieves higher accuracy with fewer computations. The square Euclidean metric is used in this manuscript to calculate distances among cities. Simulation results show that the proposed technique substantially improves many metrics for solving TSP. As a further contribution, an FPGA-based implementation is proposed; this approach has excellent scalability for many cities in the hardware implementation using FPGA as the clustering method uses the memory blocks in the FPGA. This is explained in detail in the results section.

## II. PROPOSED RECURSIVE CLUSTERING METHOD

The proposed clustering method uses a recursive clustering (RC) approach that differs from previous methods in the literature. It guarantees that a set number of cities (NOC) is divided into no more than five clusters using k-means (Algorithm A1). The size of each cluster depends on the cities' coordinates, constituting the first level of clustering. If any first-level cluster has more than five cities, it is further subdivided (lines 4 to 14 of Algorithm A1). When NOC is less than or equal to 10, two subclusters are formed; if NOC is less than or equal to 15, three subclusters are created, and so forth. This process completes the second level of clustering. If any second-level cluster still has more than five cities, it is further subdivided using the same method, generating additional levels. This continues until all clusters contain fewer than five cities.

**Algorithm A1 for clustering multiple cities**

```
Input the coordinates of all cities
1:   Clustering all cities as 5 clusters
2:   Perform SA-based TSP (Algorithm A2) for the centroids of 5
     clusters
3:   Check the number of cities (NOC) for each cluster
4:   S1: if NOC > 5
5:          k= floor (NOC÷5)
6:          if k <= 3 then
7:              If mod (NOC÷5)==0
8:                  Cluster all cities to k clusters
9:              Else
10:                 Cluster all cities to k+1 clusters
11:             Endif
12:         Else
13:             Cluster all cities to 5 clusters
14:         Endif
15:         Apply CTSP (Algorithm A3) for the centroids
              % to obtain the path among different clusters
16:         Apply CTSP for each cluster
              %find out the starting and ending of each cluster
17:      else   %NOC<=5
18:            Apply CTSP for each cluster
19:      endif
20:  Repeat S1 until the NOC of each cluster ≤ 5
```

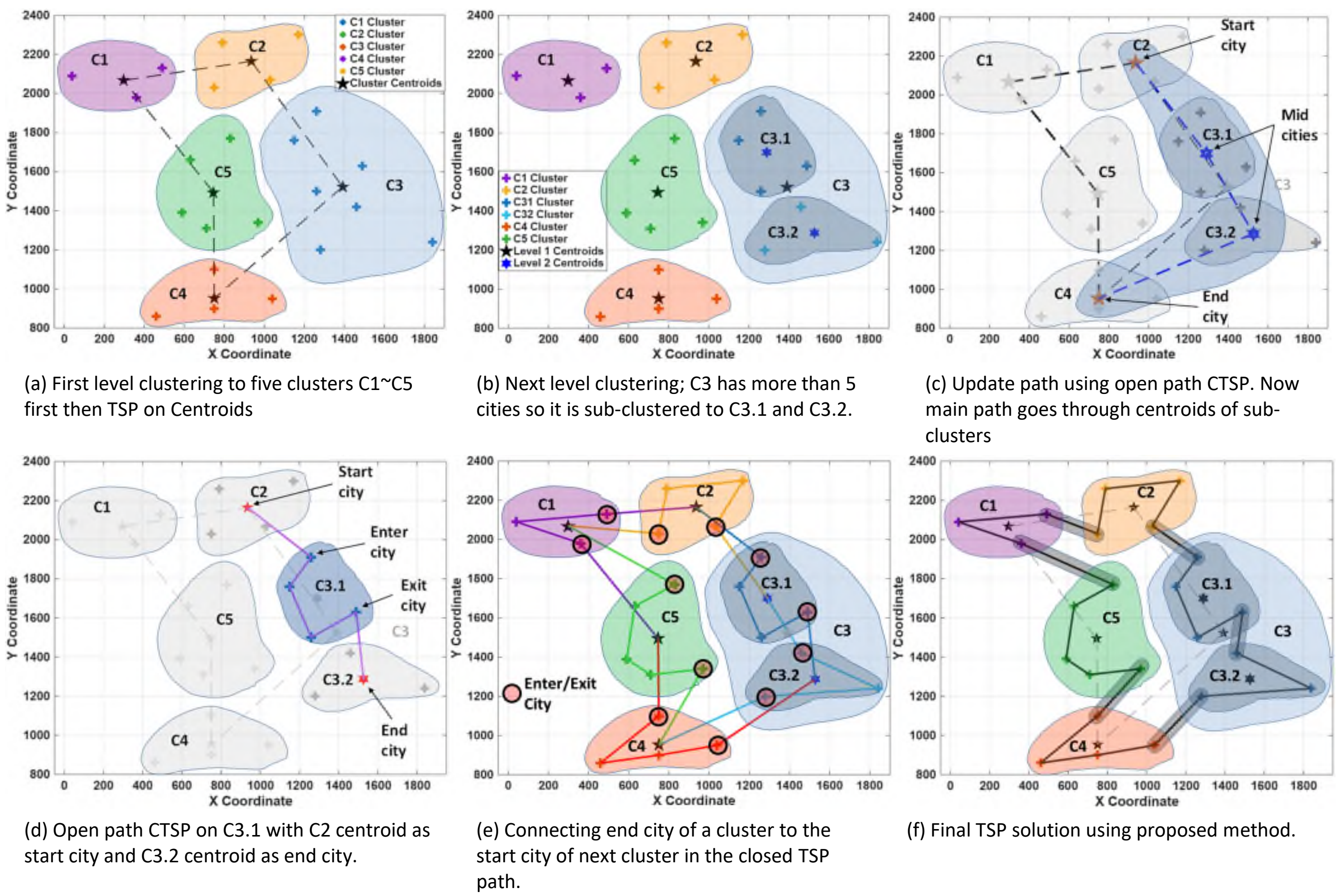

(a) First level clustering to five clusters C1~C5 first then TSP on Centroids

(b) Next level clustering; C3 has more than 5 cities so it is sub-clustered to C3.1 and C3.2.

(c) Update path using open path CTSP. Now main path goes through centroids of sub-clusters

(d) Open path CTSP on C3.1 with C2 centroid as start city and C3.2 centroid as end city.

(e) Connecting end city of a cluster to the start city of next cluster in the closed TSP path.

(f) Final TSP solution using proposed method.

**Fig. 1.** Proposed clustering method and steps of path finding.

The clustering of Algorithm A1 is in Figs. 1(a) and 1(b). In Fig. 1(a), 23 cities are first clustered into five clusters based on their coordinates using the k-means algorithm: cluster C1 has 3 cities, C2 has 4 cities, C3 has 7 cities, C4 has 4 cities, and C5 has 5 cities. Thus, level 1 clustering is formed. In this example, C3 has more than 5 cities, and therefore needs to be clustered and therefore level 2 is needed. In Fig. 1(b), C3 has been clustered into 2 sub-clusters, C3.1 and C3.2. The resulting subclusters have fewer than 5 cities: 4 in C3.1 and 3 in C3.2. If Algorithm A1 encounters any sub-cluster that contains more than 5 cities in level 2, it executes to a deeper level (level 3 in this example) and clusters those sub-clusters. This process continues till all clusters (including sub-clusters) have fewer than five cities each.

Before sub-clustering, Algorithm A1 must identify the main path solution that goes through the centroids of the five level 1 clusters. This is done by solving a TSP using a Simulated Annealing (SA)-based TSP (Algorithm A2) algorithm. The resulting TSP main closed path is shown in Fig. 1(a). Then, to incorporate any new sub-cluster into the main TSP path, the open-path Constraint TSP (CTSP) Algorithm 3 is applied to the centroids of the new sub-clusters fine-grain the main path. The input to this algorithm consists of the preceding and succeeding path nodes, which serve as the start and end cities, respectively. This is illustrated in Fig. 1(c), where C3.1 and C3.2 centroids are the mid-cities, while the centroid of C2 is the start city and the centroid of C4 is the end city. The new path is shown in Fig. 1(c) and can be compared to the original path of Fig. 1(a).

**Algorithm A2 for SA (Simulated annealing)-based TSP**

```
Input the coordinates of all cities
1:    Initialize parameters (initial temperature 1000, cooling ratio 0.9,
      Markov chain length 3000, and stopping condition res=1e-3)
2:    1. While temperature > resolution
3:          Store current best energy
4:          1.1. For each Markov chain iteration:
5:                1.1.1 Modify route: Randomly swap two cities or shift
                        segments
6:                1.1.2. Compute new energy (route cost)
7:                1.1.3. Acceptance criteria with probability
                         exp (-energy variation/temperature):
8:                       1.1.3.1. If the new route is better, accept it.
9:                       1.1.3.2. Otherwise, accept with a probability
                                  based on temperature
10:         1.2. Update best route if applicable
11:         1.3. Reduce temperature by multiplied by the cooling ratio
12:         1.4. Break condition: Stop if energy values remain unchanged
13:   2. Construct final visited coordinates
14:   3. Stop and output best route and time taken
```

Once all cities are grouped into 5 or fewer city clusters, the overall TSP path that goes through all clusters' centroids is now available. In our example, this is the path in Fig. 1(c). The next step is to identify the path connecting each cluster's cities to the main path. This is accomplished in the same way the sub-clusters were incorporated into the main path. Fig. 1(d) shows the way to find the path through the cities of cluster C3.1 and the connections of this path to the main path; the CTSP is formed by assigning the preceding cluster (C2) centroid as the start city, the succeeding cluster (C3.2) centroid as the end city, and the cities of the cluster (C3.1) as mid cities. CTSP is then executed to find a path connecting all cities.

The same procedure is performed on all clusters. The results of this process are shown in Fig. 1(e); all paths now start from a cluster centroid and end at another cluster centroid. The final step consists of connecting the cluster's exit city to the succeeding cluster's enter city. This way, all centroids are removed and all cities are now connected. Clusters' entrances and exits are shown in Fig. 1(e), and the final path is depicted in Fig. 1(f), in which the exit city of a cluster is connected to the entrance city of the following cluster in the path.

As per Algorithms C1 and C3, the TSP is solved for at most five nodes, whereas the CTSP in Algorithm A3 is solved for up to only seven nodes. This significantly accelerates the overall solution process of the TSP, particularly for large-scale instances in which the search space and the number of required iterations is large. The implemented problem consists of 7 cities, with a fixed start city and a fixed end city. Only the 5 intermediate cities are permuted, resulting in 5!=120 possible paths. On the other hand, for a general TSP with N cities and fixed start and end points, the number of permutations grows factorially ((N-2)!). For example, for N=10, the total possible paths are 8!=40,320. Therefore, because the search space for 7 cities is extremely small, near-complete exploration is achievable with a modest number of iterations.

**Algorithm A3 of CTSP (Constrained TSP) for connecting different clusters**

Input the coordinates of all one cluster from the results of Algorithm A1 as the initial path

1: 1. Insert CP=centroid of previous cluster as the starting of new path
2: 2. Insert CN=centroid of next cluster as the ending of new path
3: 3. Compute the initial distance
4: 4. Apply simulated annealing
5: 4.1. Set annealing parameters (temperature 100, cooling ratio 0.99, max iterations 1000, stopping condition res=1e-3)
6: 4.2. Generate a new route by swapping two cities
7: 4.3. Compute the new route distance
8: 4.4. Accept or reject the new route based on probability (exp (-distance variation/temperature))
9: 4.5. Update the best route if improvement is found
10: 4.6. Reduce temperature by multiplied by the cooling ratio
11: 4.7. Terminate if temperature is sufficiently low
12: 5. Return Best Route and Distance

## III. SIMULATION RESULTS

To ensure that the algorithm performs reliably regardless of system specs, all implementations in this work are carried out using Matlab 2024b on three different systems. The specifications of all three personal computers are summarized in Table I.

TABLE I
SPECIFICATIONS OF THE SYSTEMS USED IN THE SIMULATION ENVIRONMENT

| Specification | i9 system | i5 system | i7 system |
|---|---|---|---|
| Processor | i9-9980HK | i5-1235U | i7-1280P |
| Base Clock (P-cores) | 2.4 GHz | 1.3 GHz | 1.8 GHz |
| Turbo Boost (P-cores) | Up to 5.0 GHz | Up to 4.4 GHz | Up to 4.8 GHz |
| Architecture | Coffee Lake (9th Gen, 2019) | Alder Lake (12th Gen, 2022) | Alder Lake (12th Gen, 2022 |
| Cores / Threads | 8 cores / 16 threads | 10 cores / 12 threads | 14 cores / 20 threads |
| RAM | 32GB DDR4 (2666 MHz) | 16GB DDR4 (3200 MHz) | 32GB LPDDR4x (4266 MHz) |
| Storage type | PCIe 3.0 x4 SSD | PCIe 4.0 4x SSD | PCIe 4.0 4x SSD |
| Graphics | AMD Radeon Pro 5500M 8GB GDDR6 + Intel UHD Graphics 630 | Intel UHD/ Graphics | Intel Iris Xe Graphics |
| Operating System | Windows 11 Pro | Windows 10 Education | Windows 11 Home |

The proposed RC method is implemented on the same TSPLIB benchmark samples from [1] and also used in [18]. Figs. 2 and 3 present two examples of implementing the proposed Algorithms A1-A3 for the 29- and 76-city cases.

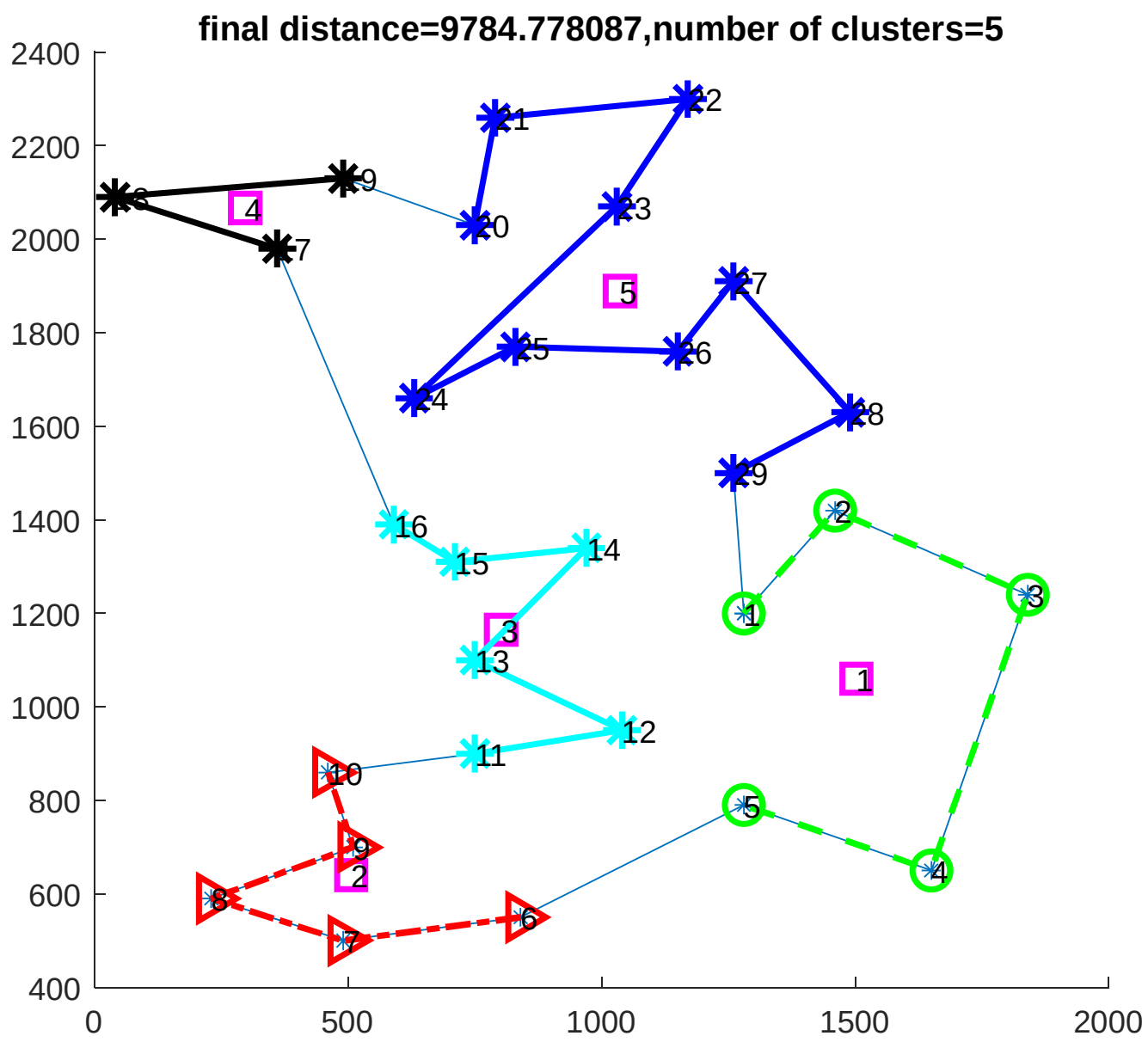

**Fig. 2.** The output for bays29.

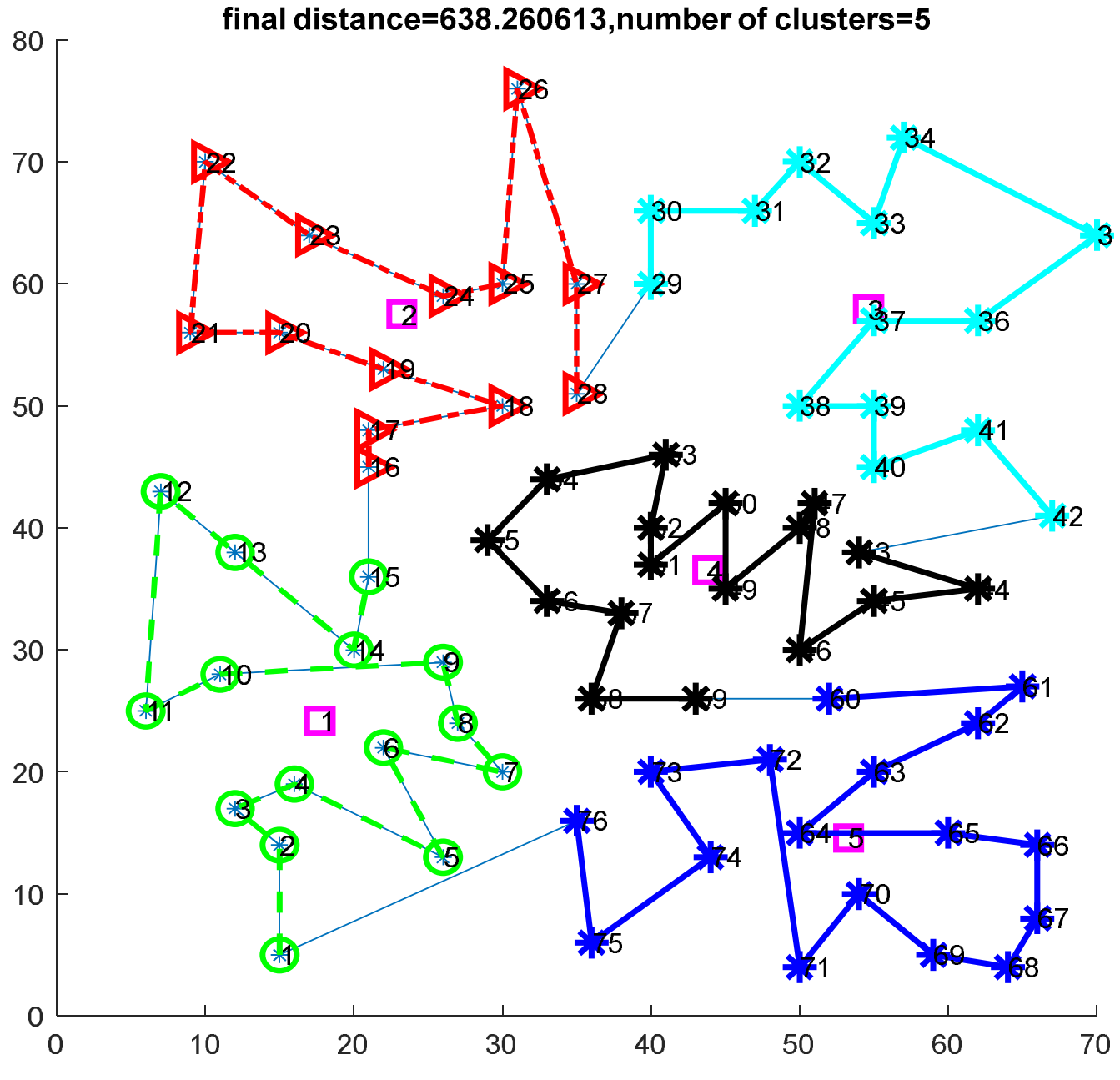

**Fig. 3**. The output for yeil76

Next, the results for the best routing distance of the proposed method (PM), RC, Simulated Annealing (SA), and [18] against the optimal distance (OD) provided by the library are presented in Figs. 4 and 5 (lower subplots). The distances for [18] are provided for city groups with up to 101 cities (ell101); the distance achieved by SA is relatively close to OD, compared with PM and [18]. Additionally, PM yields a better distance ratio than [18]. However, all these results are good for SA for the city groups with fewer cities (fewer than 101). When many cities are considered (lower subplot of Fig. 5), the PM significantly outperforms SA in the distance ratio. This can be verified by observing the PM to SA ratio in the lower subplot of Fig. 5. PM begins to demonstrate effectiveness when applied to a 200-city group (gr202) and larger. The results presented are the mean values across the three systems (Table I), i.e., the results are based on more than 300 runs for each city.

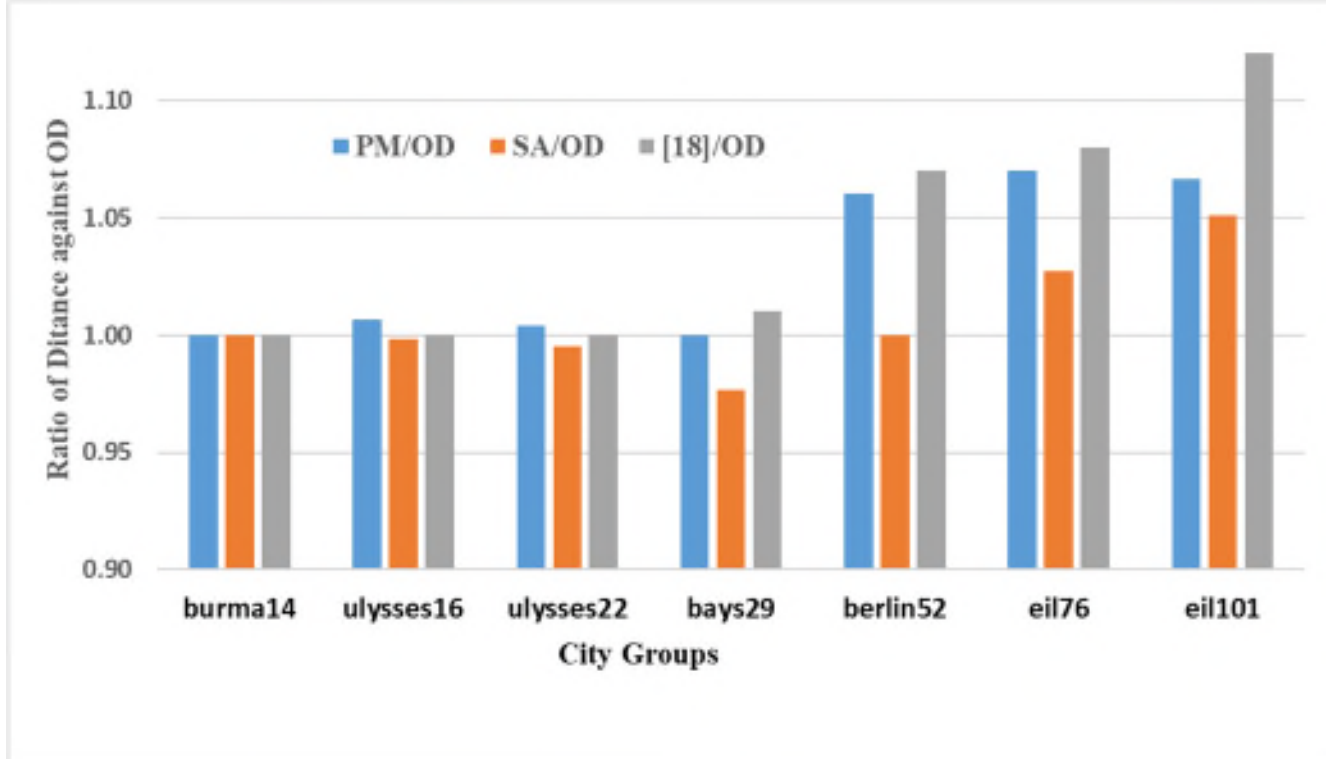

**Fig. 4.** Best Distance Ratio of PM, SA & [18] against OD for the small number of cities

The average routing distance is computed by performing 100 runs for each city group using PM, SA, and the [18] method. The ratio of each method against OD is shown in Fig. 6 and Fig. 5 (upper subplot). The city groups with a small number of cities are shown in Fig. 6, whereas those with many cities are shown in Fig. 5 (upper subplot). For the small-number-of-cities case, the SA results in the best distance ratio with OD, followed by PM and [18]. However, for many cities, PM clearly outperforms SA; the PM-to-SA ratio is shown in the upper subplot of Fig. 5.

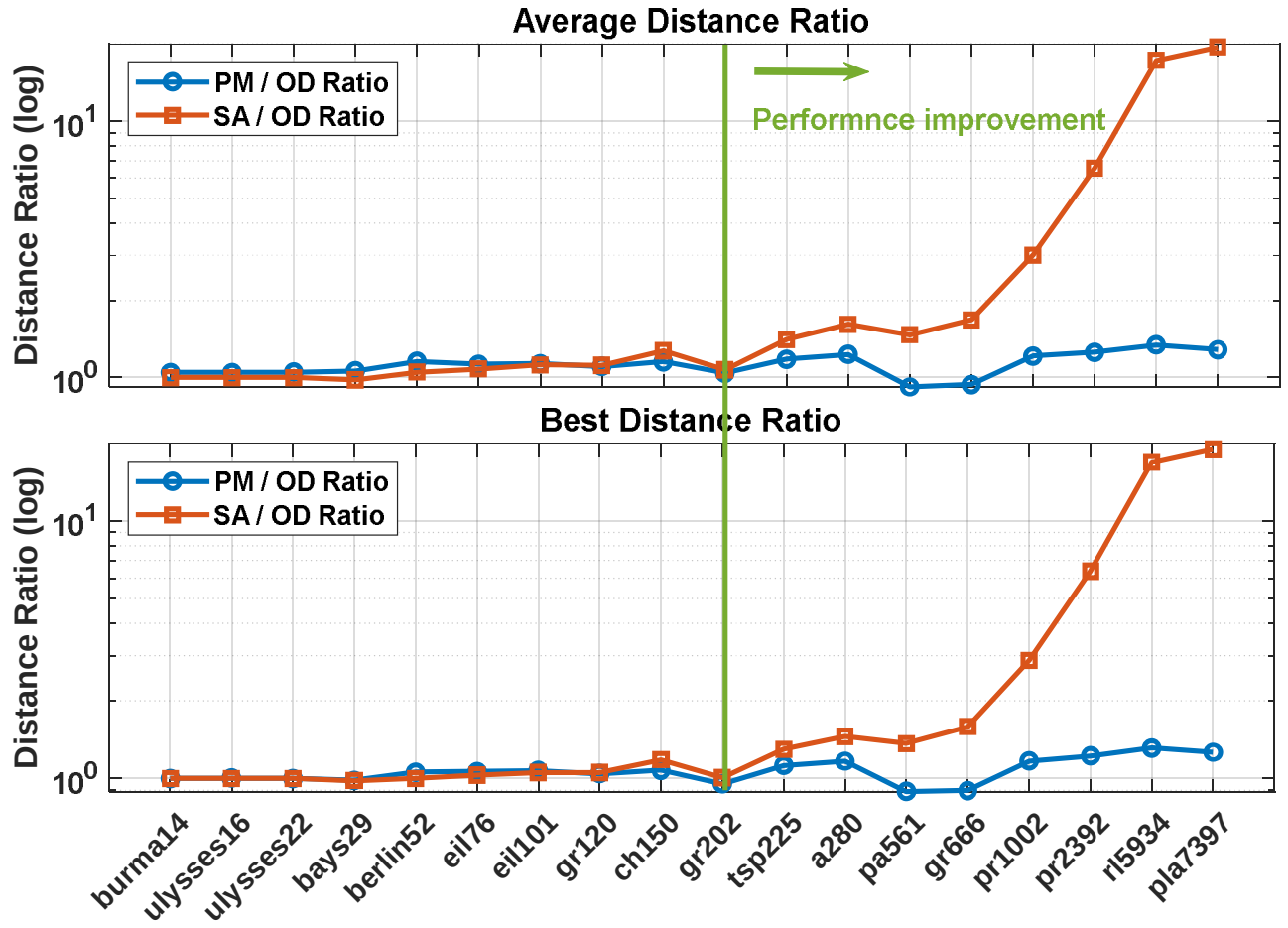

**Fig. 5.** Average (upper) and Best (lower) Distance Ratios of PM & SA against OD

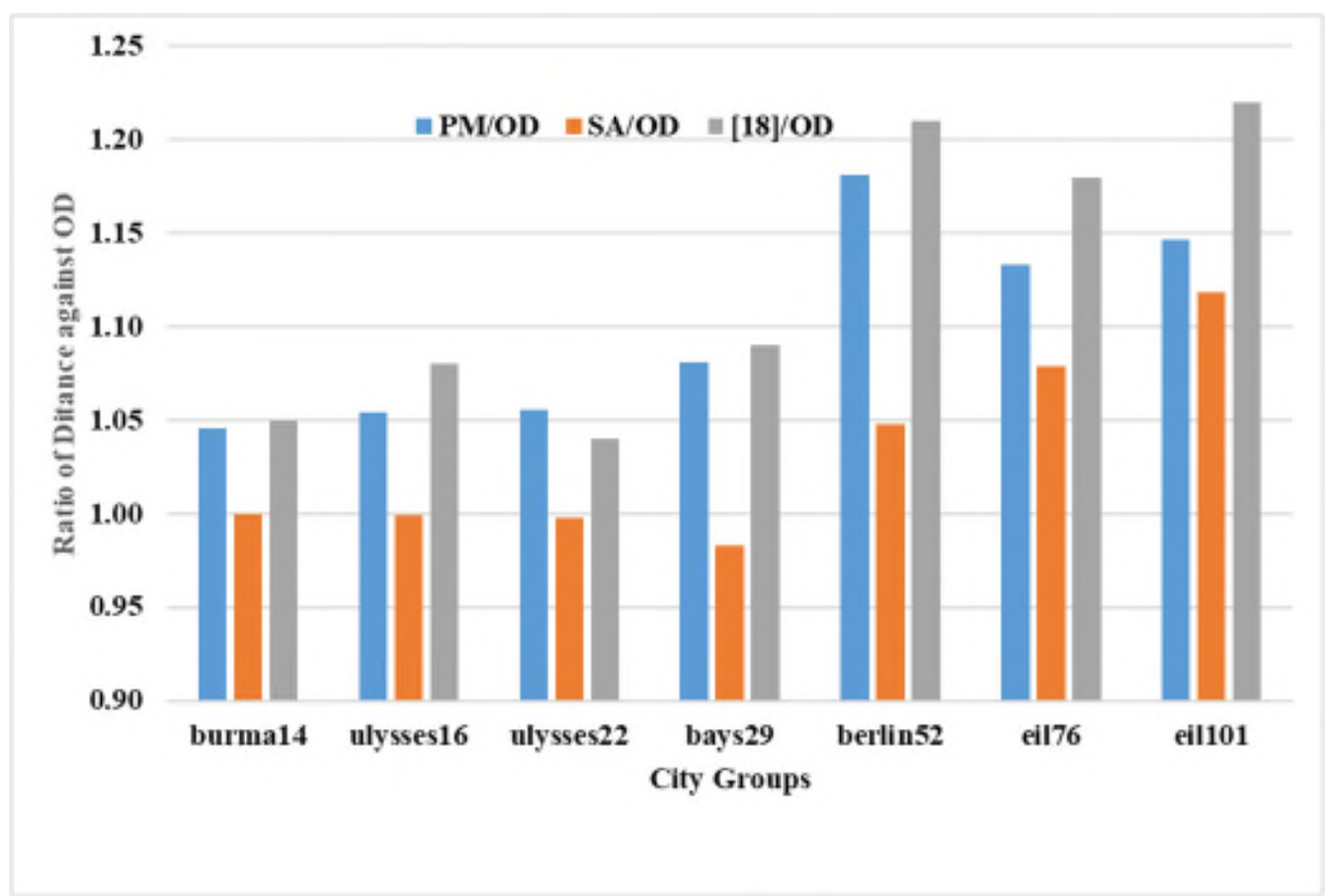

**Fig. 6** Average Distance Ratio of PM, SA & [18] against OD for the small number of cities

Fig. 7 depicts the ratio of the resulting distance from the conventional simulated annealing (SA) to that from the proposed method; as the city group size increases (gr202 and above), the ratio increases, reaching a value of 43 (i.e., the PM produces a distance 43 times shorter than SA). This is applicable to both the average and best results (Fig. 7).

Next, to leverage the proposed method's parallel implementation capability, we compare the performance of PM with that of PM's parallel implementation on the multicore systems listed in Table I. The sequential implementation is denoted as the Sequential Proposed Method (SPM), and the parallel processing implementation as the Parallel Proposed Method (PPM). Parallel processing is enabled by the proposed method, which clusters cities into independent groups. When implementing the proposed method using parallel processing, first, all cities are clustered into five clusters, and a TSP is performed on the resulting centroids. Then, we check the number of cities for each cluster. For those clusters whose NOC (number of cities) ≥ 6, parallel processing (maximum of 5 cores) is applied to further cluster each one into 2~5 sub-clusters (no more than 5) and PTSP is performed for the centroids of the resulting sub-clusters. The NOC check in each cluster is repeatedly performed, and parallel processing is used for clustering and PTSP until the NOC in each sub-cluster is at most 5. Finally, PTSP is applied to the cities in each cluster using parallel processing (up to 8 cores) to obtain the final city sequence. Note that the starting point of the PTSP for the cities of the current cluster is the centroid of the previous cluster; the ending point of the PTSP is the centroid of the next cluster.

The results are plotted in Figs. 8 and 9. SA requires approximately 2 to 4.6 times as long to obtain the solution; the parallel implementation of the proposed method for large city groups requires less than 3 times the duration of the sequential implementation, i.e., more than 3 times faster.

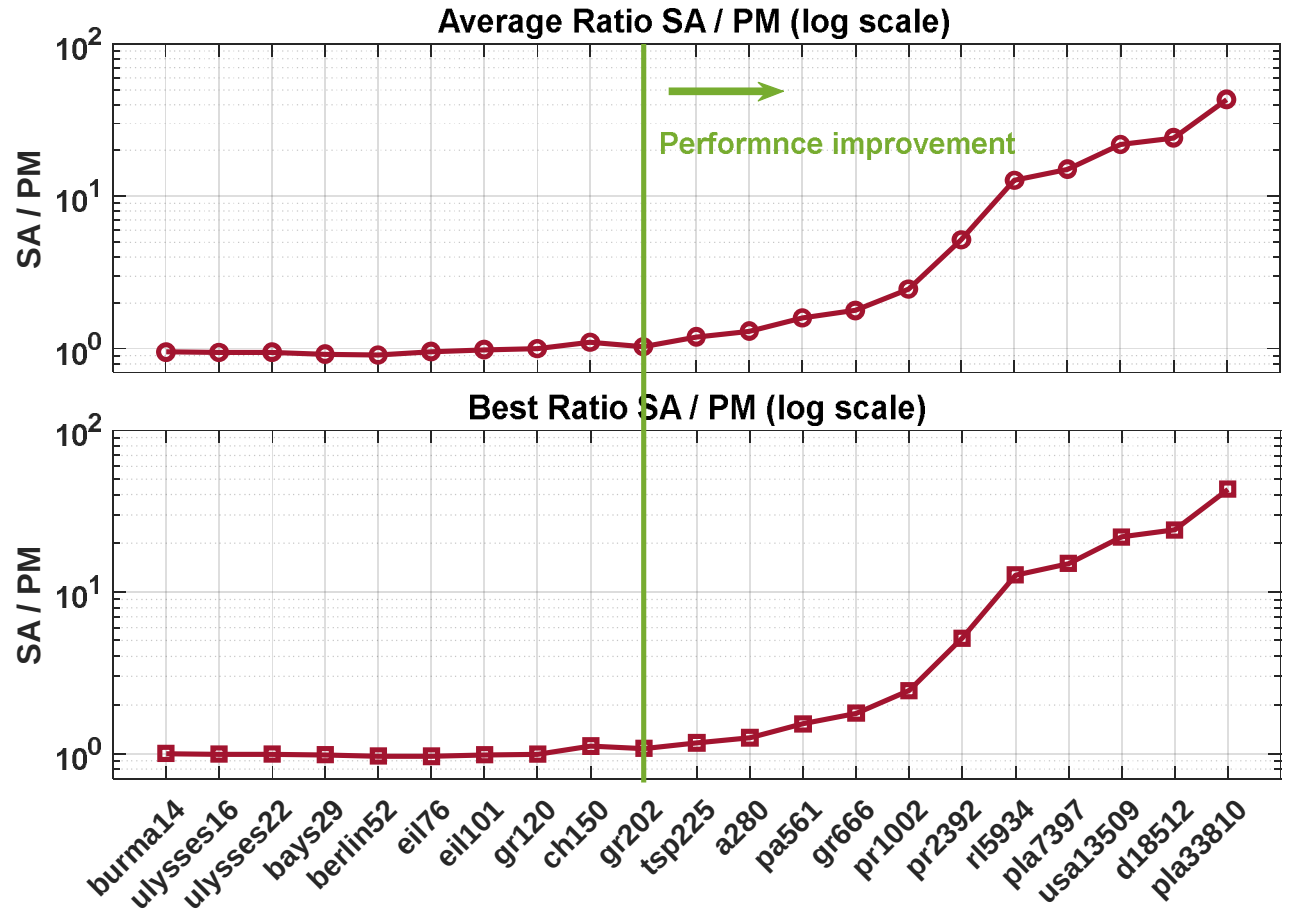

**Fig. 7.** Average (top figure) and Best (bottom figure) Distance Ratios of PM & SA against OD and against each other

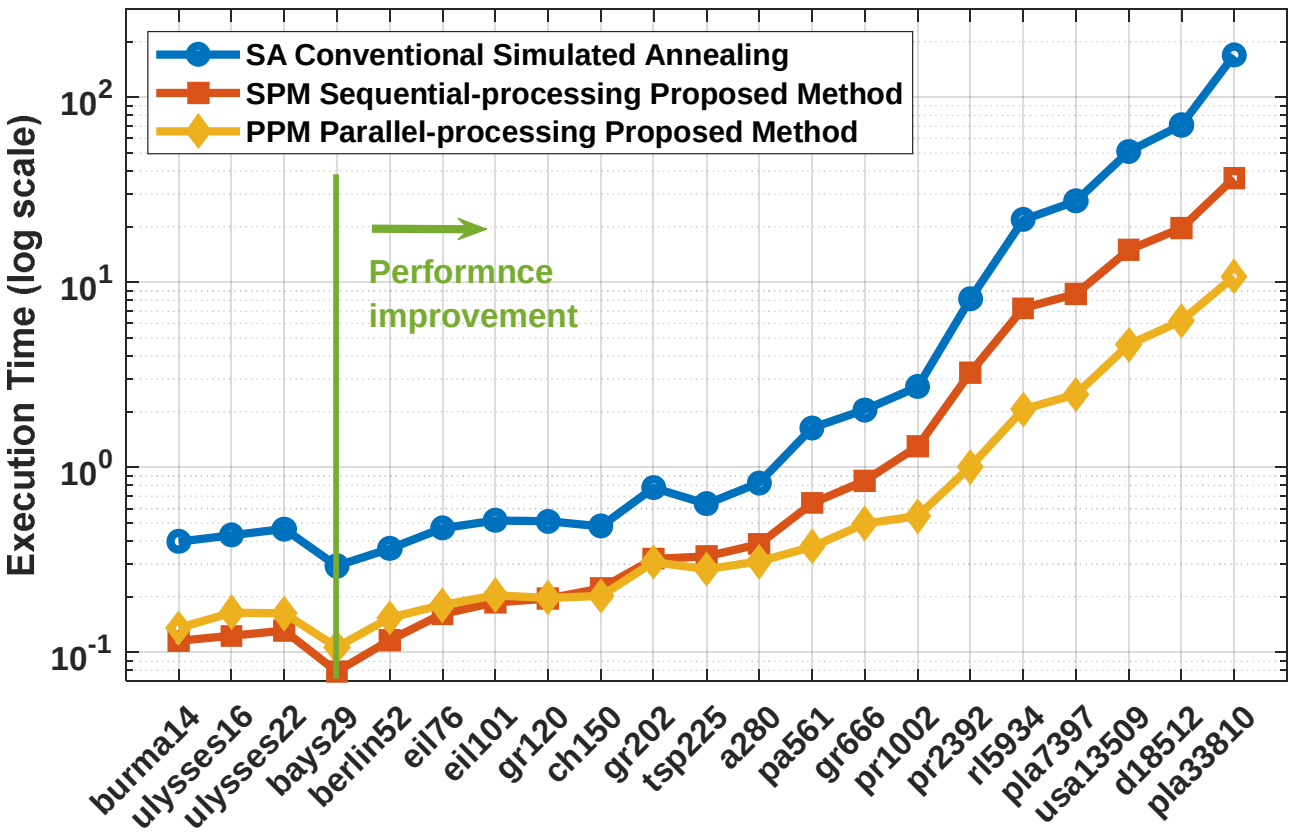

**Fig. 8.** Execution time of Average TSP results using PM (sequential), PM (parallel processing), and conventional SA.

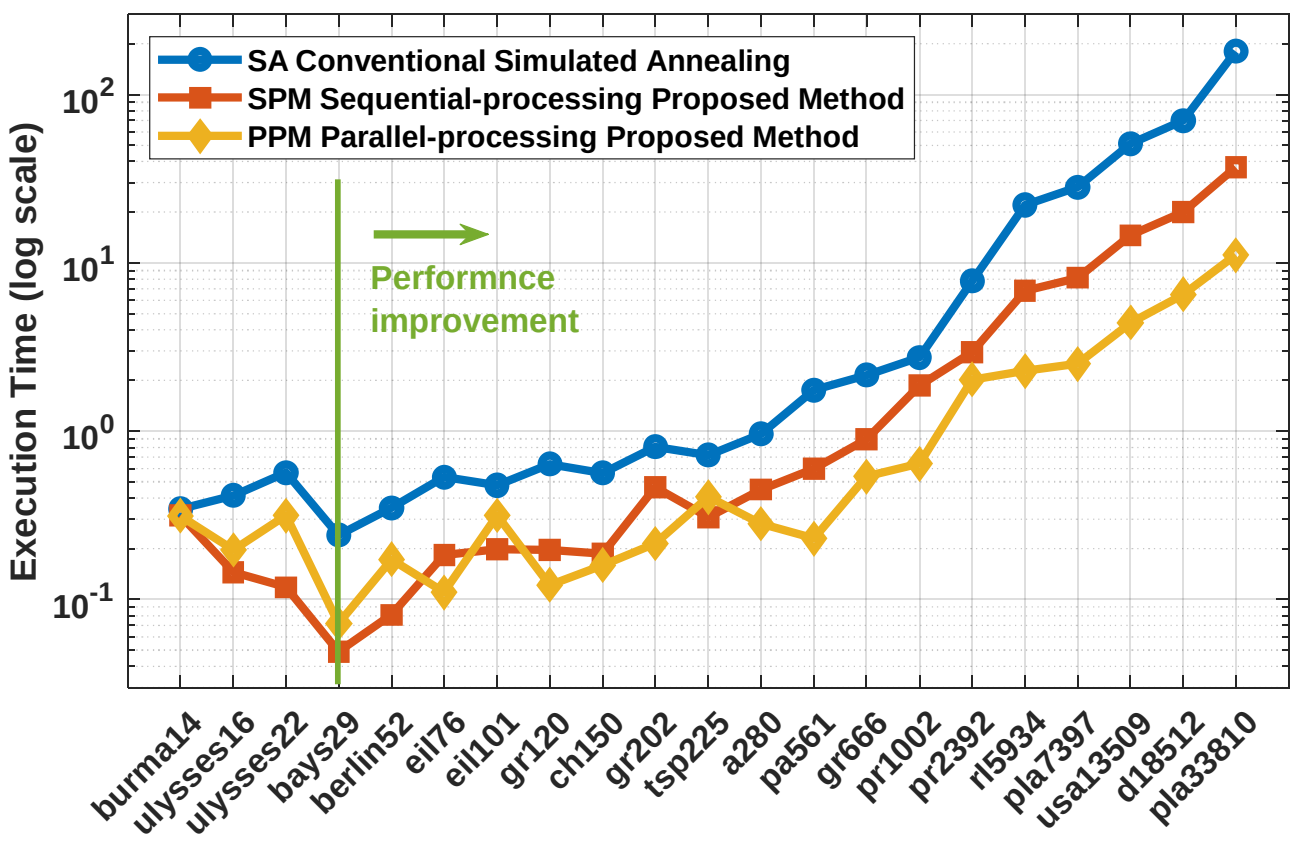

**Fig. 9.** Execution time of Best TSP results using PM (sequential), PM (parallel processing), and conventional SA.

The results of Figs. 8 and 9 were obtained using the first system in Table I (i9 system). To verify whether the proposed algorithm is system-dependent (operating system, processor architecture, RAM capacity, etc.), all methods were simulated on the three systems listed in Table I, and the results are shown in Figs. 10 and 11. The mean execution time of implementing all methods on all three systems of Table I is depicted in Fig. 12, in which the ratio of SA to PPM and the ratio of SPM to PPM are plotted.

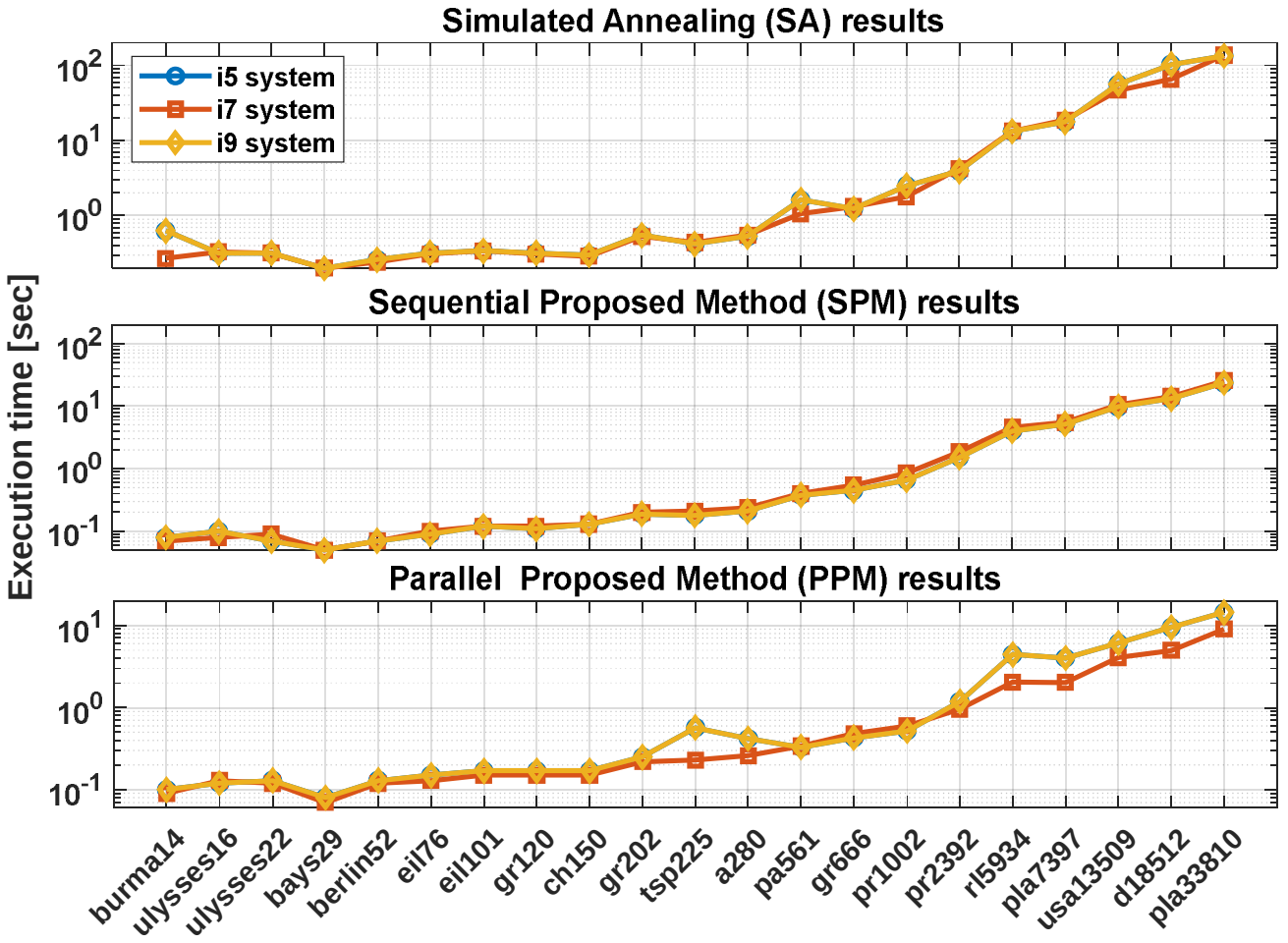

**Fig. 10.** Execution time of Average TSP results using Sequential PM (SPM), Parallel PM (PPM), and conventional Simulated Annealing (SA).

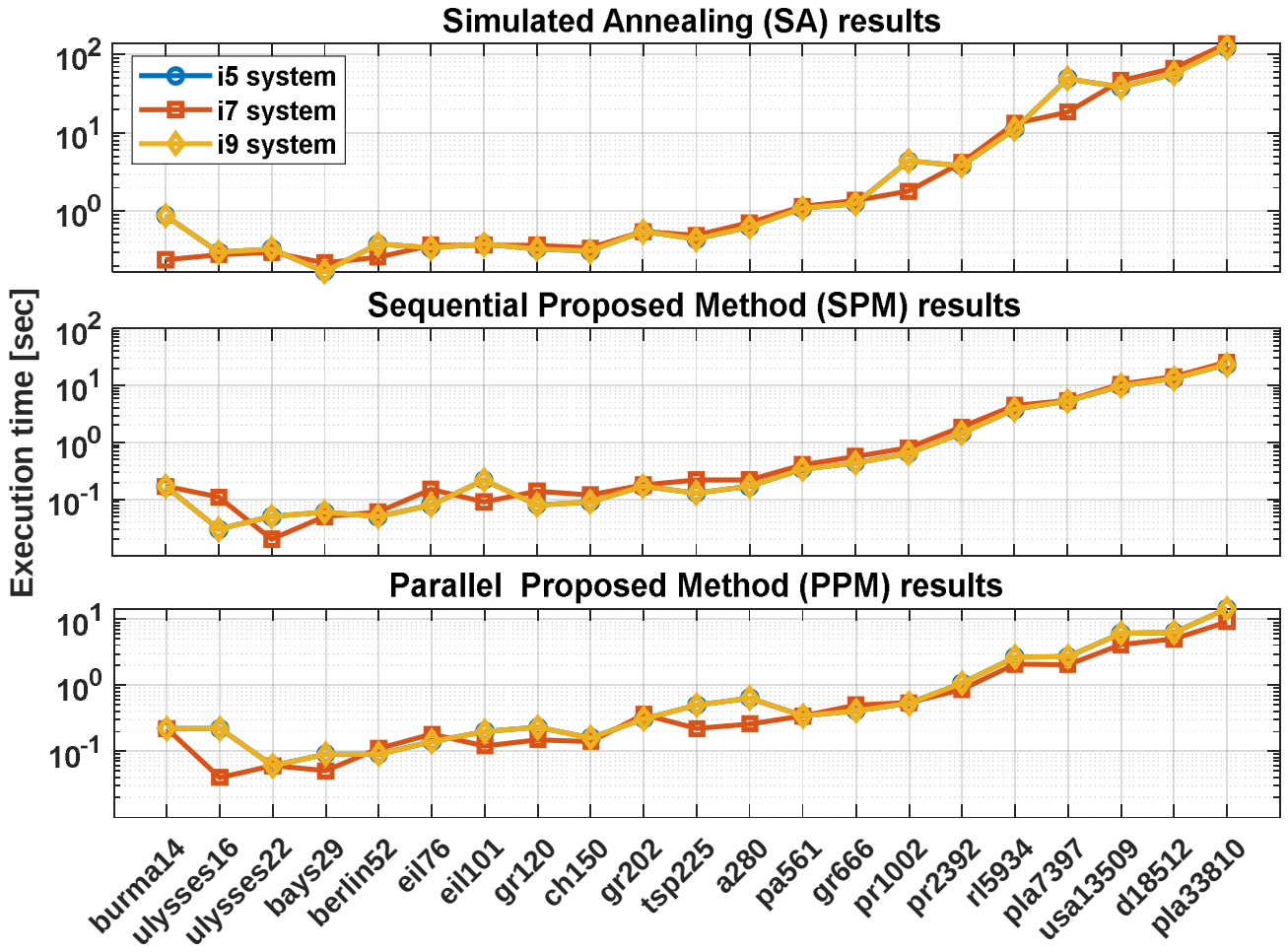

**Fig. 11.** Execution time of Average TSP results using Sequential PM (SPM), Parallel PM (PPM), and conventional Simulated Annealing (SA).

For more details, the results are also provided in Table A1 of the supplementary material, where the average maximum level denotes the average of the maximum layer reached when solving the problem across 100 runs for each city group. The layers columns denote the number of clusters that contain fewer than 6 clusters, and therefore, they do not require further clustering. For example, the city group A280 does not have any clusters in layer 1 with 5 cities or fewer, hence the number in layer 1 is 0.

Table A2 in the supplementary material plots the average and best-distance results ratios for implementing PM & SA against OD and against each other; it also shows the execution times for all SPM (sequential implementation), PPM (parallel processing implementation), and conventional SA across all city groups. Note that the conventional SA is also a sequential implementation, as it cannot be parallelized.

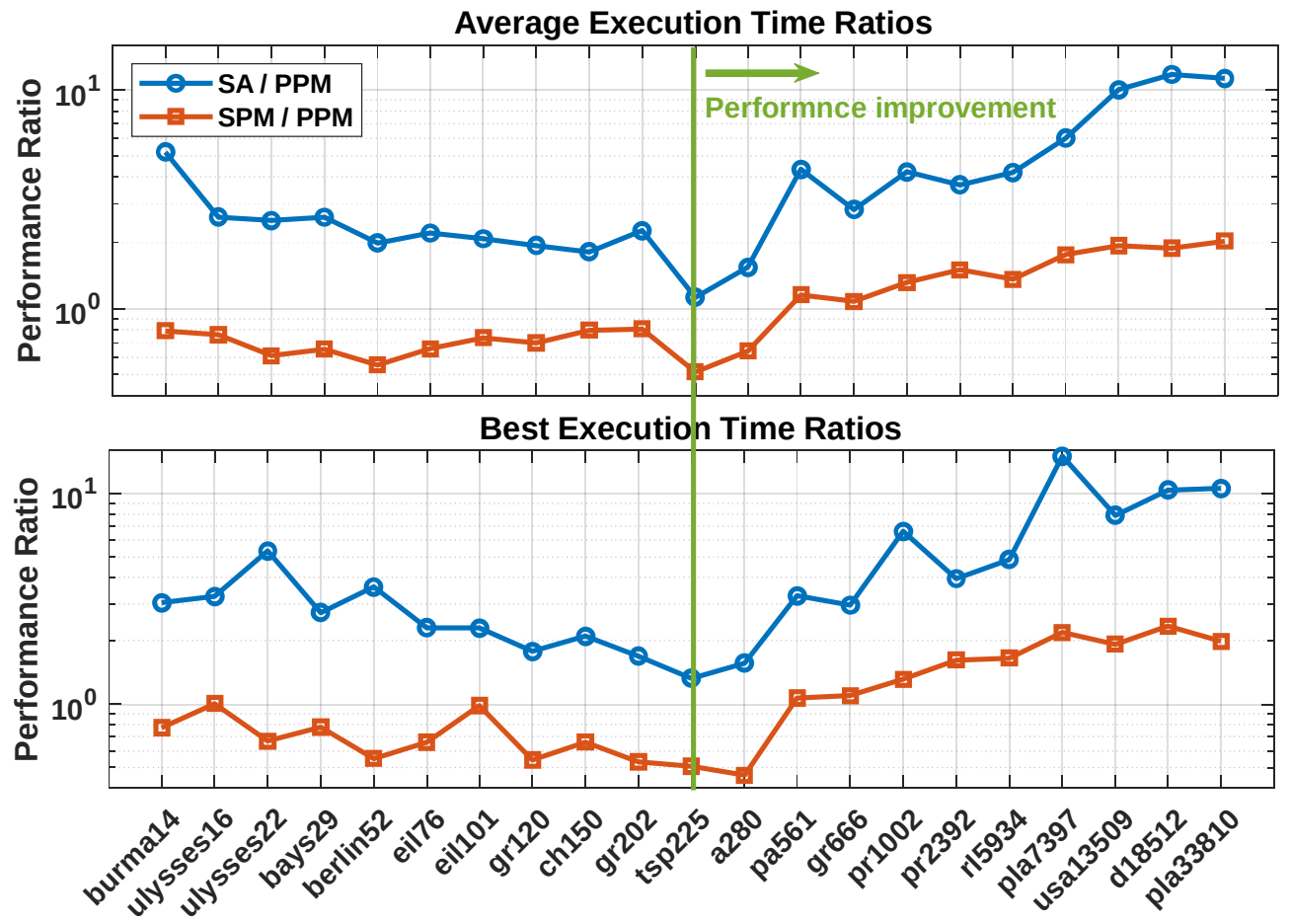

**Fig. 12.** Average (top figure) and Best (bottom figure) execution time ratios of SA to PPM, and SPM to PPM. The data is the mean of all systems (i5, i7, and i9)

## IV. Hardware Implementation

This section details the FPGA (Artix-7 (xc7a200t-2-ffg1156)) implementation of the proposed recursive K-means clustering algorithm; clusters must satisfy a strict size bound (≤5 cities), and the cluster pools are stored in memory (in this case, RAMB36-based BRAMs) for deterministic access. In addition, the implementation of the constrained TSP and SA completes the routing.

In the K-means algorithm, the distance between city coordinates $(x, y)$ can be evaluated using several commonly used metrics, as summarized in Table II.

TABLE II
DISTANCE CALCULATION FUNCTION METHODS

| Method | Formula |
|---|---|
| Euclidean | $\sqrt{\Delta x^2 + \Delta y^2}$ |
| Squared Euclidean | $\Delta x^2 + \Delta y^2$ |
| Manhattan | $\lvert\Delta x\rvert + \lvert\Delta y\rvert$ |
| Octile | $\max(\Delta x, \Delta y) + 0.414 \cdot \min(\Delta x, \Delta y)$ |

To evaluate the performance of the different distance methods in hardware, their implementation has been pursued on the Artix-7 FPGA (xc7a200t-2-ffg1156). The results for these methods are reported in Table III; the Squared Euclidean distance requires the lowest number of LUTs and registers, while finding the shortest combinational path delay. Therefore, although the Euclidean method yields a marginally shorter tour length (Figure 15), the proposed K-means hardware implementation uses the Squared Euclidean distance, because it offers the most favorable trade-off between clustering performance and hardware efficiency for an FPGA-based implementation.

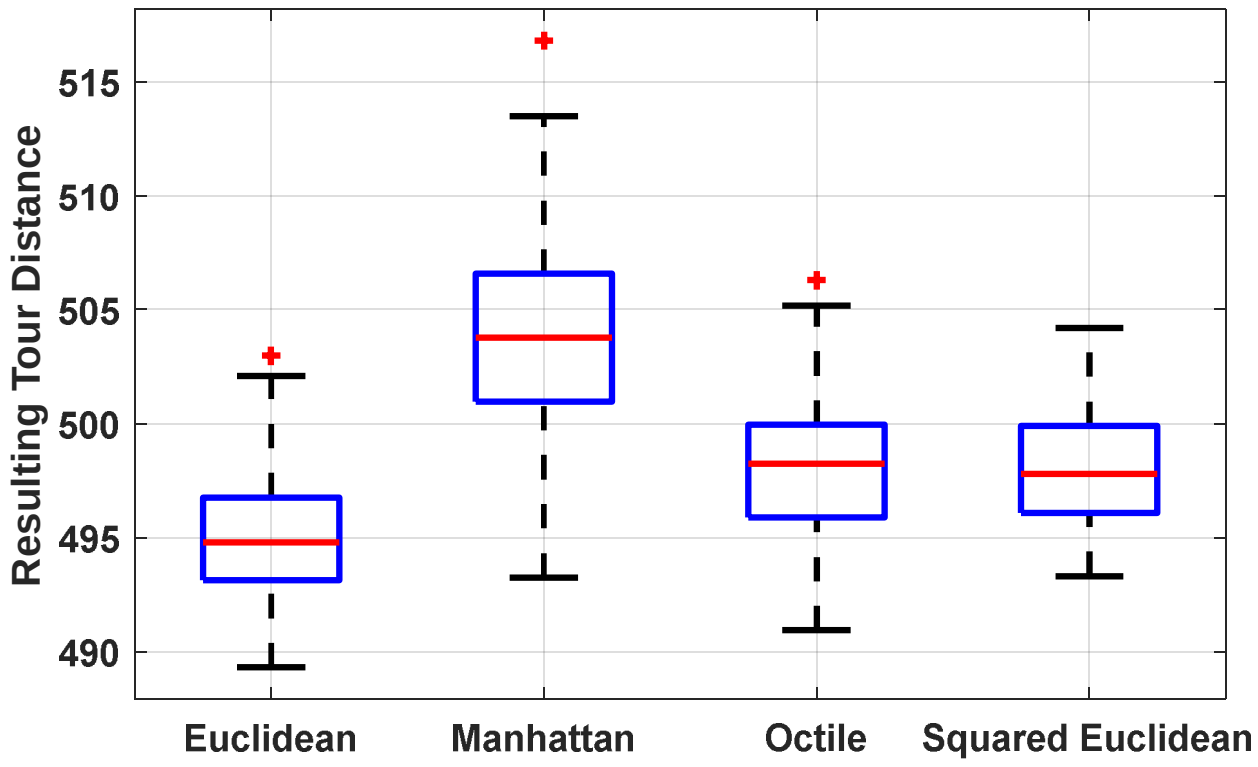

**Fig. 15.** Comparison of distance calculation methods using SA-based TSP on 202 cities, repeated 100 times, based on simulation results.

TABLE III
HARDWARE UTILIZATION OF DISTANCE CALCULATION METHODS

| Method | No. of Slice LUTs | Shortest combinational path delay (ns) | No. of Registers | No. of DSP Blocks |
|---|---|---|---|---|
| Euclidean | 204 | 10.836 | 57 | 2 |
| Octile | 139 | 12.618 | 0 | 1 |
| Squared Euclidean | 32 | 7.068 | 0 | 2 |

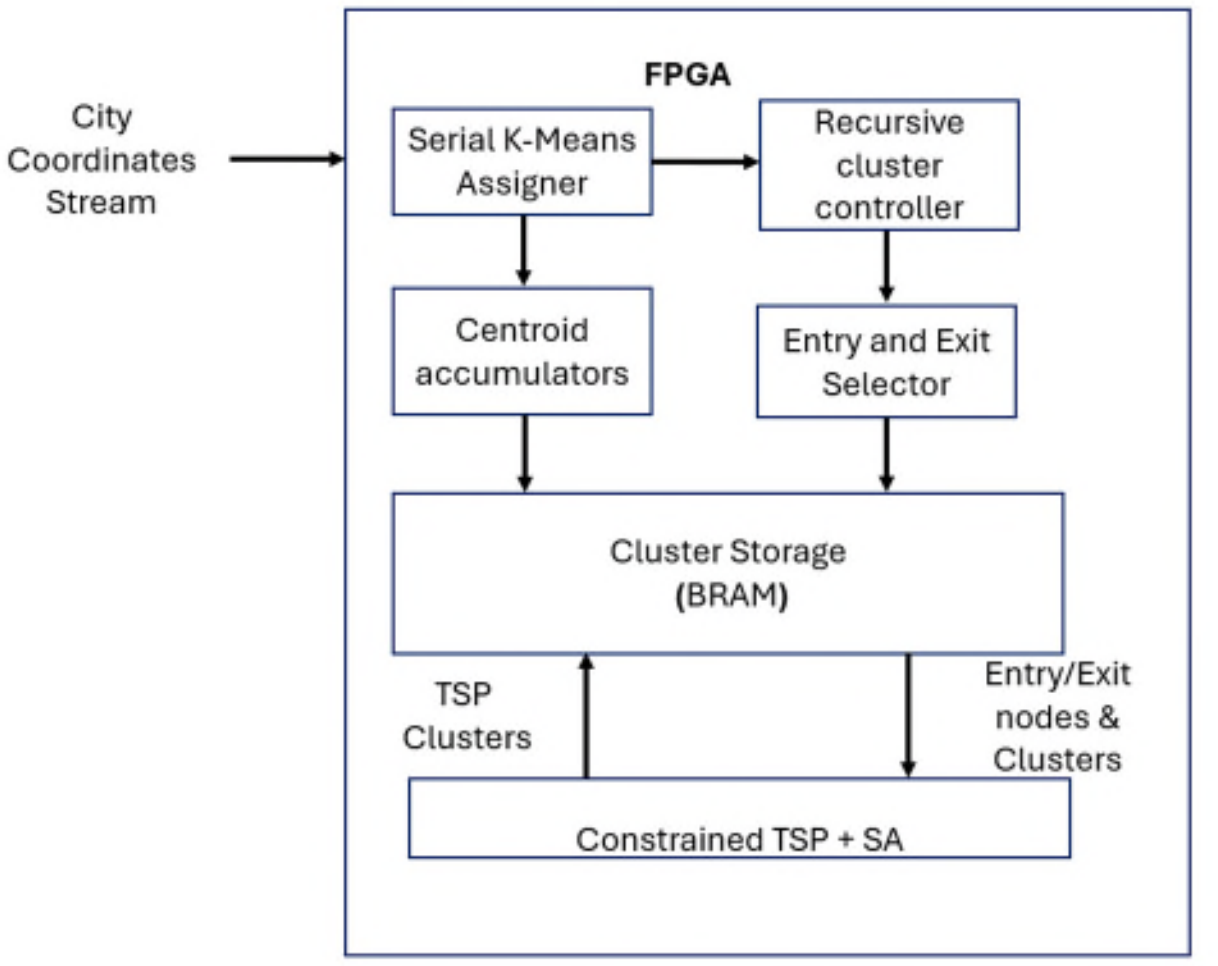

**Fig. 16.** Block diagram of the proposed FPGA-based recursive clustering and constrained TSP with SA.

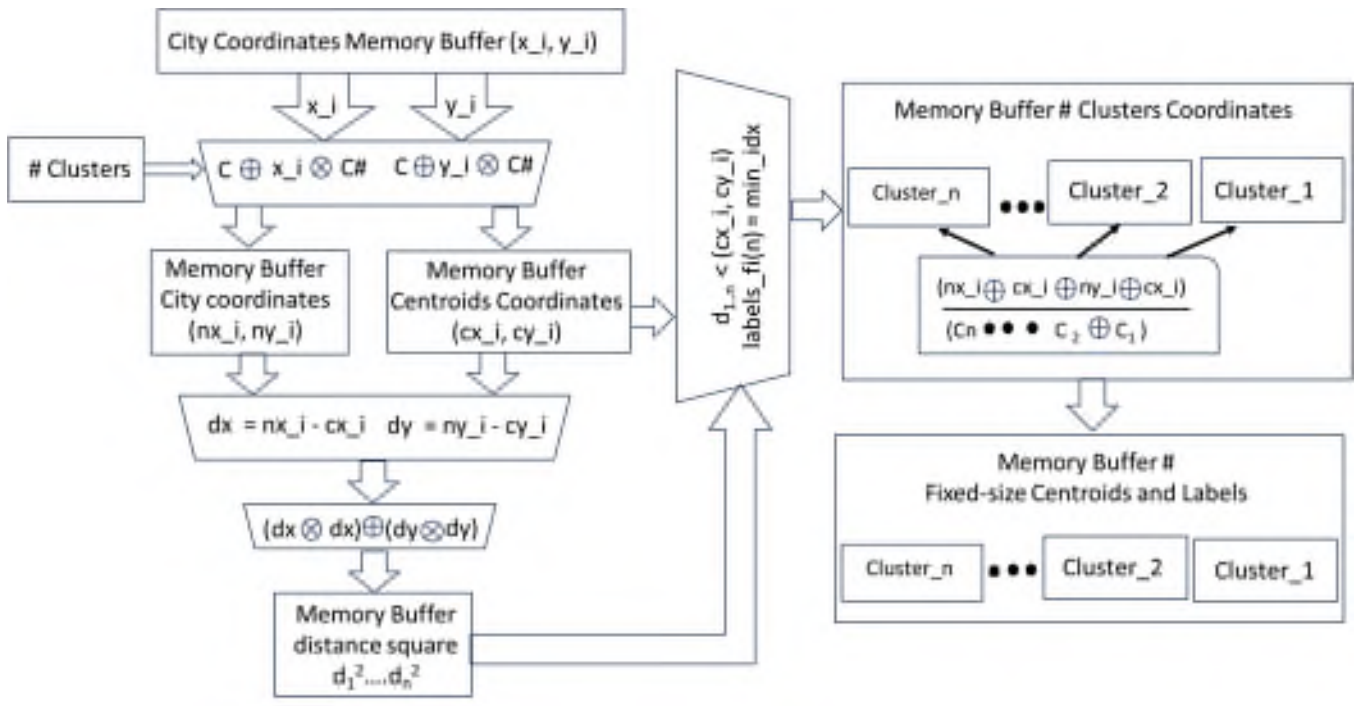

**Fig. 17.** K-means clustering FPGA implementation block diagram.

Figure 16 shows the block diagram of the FPGA (Xilinx Artix-7 (xc7a200t-2-ffg1156)) implementation of the K-means recursive clustering algorithm and constrained TSP with SA. This design handles deterministic preprocessing tasks, such as clustering, centroid computation, and entry-exit node selection, as well as the constrained TSP (CTSP) and SA performed on each cluster. In this setup, city coordinates are streamed into the FPGA and clustered via a recursive K-means method (Figure 17) until each cluster contains no more than five cities. These final clusters form manageable sub-problems for efficient CTSP processing. External inputs provide the initial centroids; for each city, the squared Euclidean distance to every centroid is sequentially calculated using a single arithmetic data path, so reducing hardware utilization (1% of the available hardware, Table IV) and keeping it constant regardless of the data set. The city is assigned to the nearest centroid, and cluster statistics, such as coordinate sums and city counts, are updated to the RAMB. Each cluster has a dedicated RAMB36 block capable of storing up to 1,152 city coordinates as 32-bit words (16 bits each for x and y). A total of 365 RAMB blocks are available in this FPGA; therefore, accommodating a maximum of 420480 city coordinates. After the initial clustering, clusters with more than five cities are recursively subdivided by reapplying K-means to their data stored in BRAM. This continues hierarchically until all clusters satisfy the condition that they have fewer than or equal to five cities. For each final cluster, the FPGA determines the entry and exit nodes for constrained routing. The entry node is chosen as the closest city to the preceding cluster's centroid, while the exit node is selected in relation to the next cluster's centroid.

These calculations use straightforward distance comparisons and BRAM reads, making them suitable for hardware implementation. Table IV displays the hardware utilization for routing 5760 cities. The LUT and registers are used for distance determination, whereas RAMB blocks store city coordinates. As previously explained, hardware for distance calculation remains constant regardless of the number of cities, since the process is sequential. Meanwhile, the number of clusters increases with the number of cities to be routed and depends on RAMB availability. Thus, the scalability of the proposed method depends on the availability of memory in the given FPGA.

TABLE IV
FPGA HARDWARE UTILIZATION OF THE PROPOSED METHOD

| | |
|---|---|
| Number of bonded IOBs: | 195 out of 500 39% |
| Number of RAMB36E1/FIFO36E1s | 5 out of 365 1% |
| Number of Slice LUTs | 45 out of 134,600 1% |
| Number of Slice Registers | 68 out of 269,200 1% |

The implementation of CTSP and SA requires 47055 LUTs (out of 134,600 i.e. 34%). At most, three CTSPs and SAs can be implemented concurrently.

Recent advances have demonstrated notable progress in solving the TSP using annealing-based and Ising-model accelerators. A fully integrated annealing processor for autonomous navigation that delivers high throughput by tightly integrating hardware of the annealing process has been introduced in [27]. Similarly, [25] has developed an approximate parallel annealing Ising machine aimed at embedded systems, emphasizing scalability through parallelism. A scalable in-memory clustered annealer that utilizes device-level stochasticity for TSP problems has been presented in [28]. A clustering-based TSP decomposition using an Ising solver has been presented in [18]; meanwhile, [29] and [30] have investigated in-memory annealing units and neuro-Ising hybrid solvers, respectively, achieving remarkable energy efficiency and scalability. The method presented in [26] employs a scalable, decoupled clustering annealing processor to solve large TSPs. This method is based on a decoupled hierarchical clustering algorithm, which improves both convergence and scalability by eliminating interconnections between adjacent clusters.

TABLE V
COMPARISON OF THE PROPOSED METHOD WITH THE PREVIOUS METHODS

| Method | Hardware platform | # of Cities | Clock (MHz) | Recursive hierarchy | Scalability | Solver flexibility | Annealing | Power (W) |
|---|---|---|---|---|---|---|---|---|
| [27] | 40nm CMOS | 32768 cities @ 8 chips | 200 | No | Yes Multichip | Low | On chip | 0.11@1chip |
| [25] | 28nm CMOS | 22 cities | 200 | No | No | Low | On chip | 0.041 |
| [28] | 14nm FinFET | 1060 cities | N/A | No | Yes Multichip | Low | On chip | 0.006 |
| [18] | CPU | 101 cities | 2600 | No | N/A | Medium | On chip | ~150 |
| [29] | 1T1R RRAM | 10 cities | N/A | No | No | Low | On chip | 0.045 |
| [30] | CPU+4GPU | 5934 cities | N/A | No | N/A | Low | On chip | >1000 |
| [26] | FPGA | 11849 cities @ 1 Chip 85900 cities @ 6 Chips | 125 | No | Yes Multichip | Low | On chip | ~19.9 |
| **Proposed method** | **FPGA** | **397440 cities @ 1 chip** | **393** | **Yes** | **Yes BRAM** | **High** | **On-chip** | **~20** |

Unlike previous methods, this study focuses on a deterministic hierarchical decomposition approach that uses FPGA-based recursive K-means clustering. As explained earlier, the scalability of this method depends on the availability of BRAM blocks on the FPGA that store the city coordinates generated using the recursive clustering method. The specific Artix-7 FPGA accommodates up to 420480 cities, the highest reported in the current literature.

While the proposed FPGA-based recursive clustering framework enables scalable preprocessing for large traveling salesman problems, it is constrained by on-chip memory hardware. Despite this limitation, the proposed approach offers a practical and scalable solution to large, computationally expensive routing problems. Table V compares the proposed method with other existing approaches; using a single FPGA, the proposed method can handle routing for more cities than the other methods. In addition, the proposed method shows better scalability and operating frequency.

## V. CONCLUSION

This work has introduced a recursive clustering approach and a routing method using CTSP and SA to solve the transportation problem. The goal is to identify the shortest path that covers all cities within a specific region and to develop a scalable hardware implementation. The proposed methodology involves dividing cities into clusters at each step, continuing until fewer than five cities remain. The coordinates of the clustered cities are stored in the FPGA's RAMB for routing via CTSP and SA.

The proposed algorithm consists of seven cities with fixed start and end points. Only the five intermediate cities are permuted, resulting in 5! = 120 possible paths. For comparison, a 10-city problem requires evaluating 8! = 40,320 paths, meaning the 7-city search space represents only about 0.3% of the 10-city case. This relatively small search space allows a near-complete exploration with a modest number of iterations.

FORMIA For an efficient hardware implementation, different distance calculation methods were evaluated. Although the Euclidean metric provides slightly shorter tours, the Squared Euclidean distance was selected due to its lower hardware and reduced combinational delay, so providing a better trade-off between routing performance and hardware efficiency.

This approximation-based approach has ensured that hardware scaling depends on memory and is independent of hardware, thereby enhancing the number of cities that can be routed.. The proposed implementation operates at a clock frequency of 393 MHz and, through hierarchical clustering and efficient BRAM utilization, it can support up to 397,440 cities on a single Artix-7 FPGA. Moreover, the hardware for CTSP and SA suggests that simultaneous routing of three clusters is feasible on the FPGA

Compared with prior work, the proposed method can handle more cities on a single FPGA while achieving a higher operating frequency. It has also shown improved accuracy, speed, and cost efficiency. These results indicate that combining recursive clustering with hardware-accelerated simulated annealing provides an effective and scalable framework for FPGA-based solutions to large-scale routing and transportation optimization problems.

# Supplementary material

Table A1 Resulting layers of the proposed method implementation. Last two columns are the resulting distance proposed method and optimal distance [4].

| City group | 100 Runs | Average Maxmum level | Number of clusters created in each layer in the proposed method | | | | | | | | | | | Distance | |
|---|---|---|---|---|---|---|---|---|---|---|---|---|---|---|---|
| | | | layer 1 | layer 2 | layer 3 | layer 4 | layer 5 | layer 6 | layer 7 | layer 8 | layer 9 | layer 10 | layer 11 | Proposed method | Optimal |
| burma14 | Average distance | 1.31 | 4.69 | 0.62 | | | | | | | | | | 32.244 | 30.88 |
| | Shortest distance | 2 | 4 | 2 | | | | | | | | | | 30.879 | 30.88 |
| ulysses16 | Average distance | 2.32 | 4.03 | 1.54 | 0.73 | 0.14 | | | | | | | | 77.897 | 74.11 |
| | Shortest distance | 2 | 3 | 4 | | | | | | | | | | 74.460 | 74.11 |
| ulysses22 | Average distance | 2.84 | 3.03 | 3.38 | 1.66 | 0.08 | | | | | | | | 79.649 | 75.67 |
| | Shortest distance | 3 | 3 | 3 | 2 | | | | | | | | | 75.954 | 75.67 |
| bays29 | Average distance | 2.15 | 2.16 | 5.59 | 0.3 | | | | | | | | | 9862.960 | 9291.35 |
| | Shortest distance | 2 | 2 | 6 | | | | | | | | | | 9112.542 | 9291.35 |
| berlin52 | Average distance | 3.81 | 1.14 | 7.77 | 5.92 | 1.71 | 0.3 | | | | | | | 8786.987 | 7544.37 |
| | Shortest distance | 3 | 1 | 8 | 6 | | | | | | | | | 8115.499 | 7544.37 |
| eil76 | Average distance | 3.26 | 0 | 13.04 | 7.94 | 0.51 | 0.02 | | | | | | | 616.452 | 545.39 |
| | Shortest distance | 3 | 0 | 14 | 8 | | | | | | | | | 579.037 | 545.39 |
| eil101 | Average distance | 3.99 | 0 | 14.45 | 12.38 | 2.49 | 0.34 | | | | | | | 730.271 | 642.31 |
| | Shortest distance | 4 | 0 | 12 | 14 | 6 | | | | | | | | 694.010 | 642.31 |
| gr120 | Average distance | 3.97 | 0.01 | 14.55 | 16.81 | 3.5 | 0.24 | | | | | | | 1855.842 | 1666.51 |
| | Shortest distance | 4 | 0 | 16 | 12 | 8 | | | | | | | | 1710.670 | 1666.51 |
| ch150 | Average distance | 4.12 | 0 | 11.01 | 25.78 | 5.86 | 0.27 | 0.02 | | | | | | 7582.972 | 6532.28 |
| | Shortest distance | 4 | 0 | 13 | 25 | 2 | | | | | | | | 7006.742 | 6532.28 |
| gr202 | Average distance | 5.78 | 1.04 | 5.82 | 31.39 | 19.8 | 3.33 | 1.27 | 0.22 | | | | | 574.686 | 550.00 |
| | Shortest distance | 5 | 1 | 7 | 28 | 17 | 6 | | | | | | | 534.275 | 550.00 |
| tsp225 | Average distance | 4.29 | 0 | 3.81 | 43.17 | 16.01 | 0.68 | | | | | | | 4603.335 | 3859.00 |
| | Shortest distance | 4 | 0 | 6 | 41 | 14 | | | | | | | | 4360.340 | 3859.00 |
| a280 | Average distance | 4.36 | 0 | 1 | 52.15 | 25.87 | 0.94 | | | | | | | 3196.136 | 2586.77 |
| | Shortest distance | 4 | 0 | 1 | 55 | 24 | | | | | | | | 3004.577 | 2586.77 |
| pa561 | Average distance | 5.04 | 0 | 0.03 | 72.14 | 77.62 | 7.29 | 0.1 | | | | | | 17848.752 | 19330.79 |
| | Shortest distance | 5 | 0 | 0 | 74 | 77 | 8 | | | | | | | 17226.059 | 19330.79 |
| gr666 | Average distance | 7.11 | 0 | 2.63 | 49.3 | 92.62 | 48.89 | 8.68 | 1.66 | 0.55 | 0.02 | | | 3718.404 | 3952.54 |
| | Shortest distance | 7 | 0 | 3 | 54 | 79 | 52 | 11 | 2 | | | | | 3520.045 | 3952.54 |
| pr1002 | Average distance | 5.97 | 0 | 0 | 35.89 | 171.07 | 83.25 | 6.01 | 0.1 | | | | | 313631.453 | 259066.66 |
| | Shortest distance | 6 | 0 | 0 | 34 | 174 | 77 | 10 | | | | | | 302479.981 | 259066.66 |
| pr2392 | Average distance | 6.34 | 0 | 0 | 3.06 | 318.14 | 321.37 | 44.95 | 0.84 | | | | | 473280.887 | 378032.00 |
| | Shortest distance | 6 | 0 | 0 | 4 | 321 | 311 | 48 | | | | | | 456172.596 | 378032.00 |
| rl5934 | Average distance | 7.38 | 0 | 0 | 0.64 | 121.02 | 1052.76 | 498.46 | 36.3 | 0.96 | | | | 739669.348 | 556045.00 |
| | Shortest distance | 7 | 0 | 0 | 1 | 123 | 1054 | 492 | 20 | | | | | 722300.521 | 556045.00 |
| pla7397 | Average distance | 7.37 | 0 | 0 | 2.49 | 137.64 | 1019.98 | 840.59 | 100.07 | 0.98 | | | | 29678149.607 | 23260728.00 |
| | Shortest distance | 7 | 0 | 0 | 1 | 125 | 1125 | 782 | 84 | | | | | 29033764.056 | 23260728.00 |
| usa13509 | Average distance | 9.26 | 0 | 0 | 0.14 | 62.74 | 1242.51 | 1871.41 | 737.25 | 135.2 | 14.66 | 0.59 | 0.02 | 25358286.89 | - |
| | Shortest distance | 9 | 0 | 0 | 0 | 47 | 1301 | 1832 | 758 | 113 | 16 | | | 24932415.93 | - |
| d18512 | Average distance | 8.81 | 0 | 0 | 0.02 | 1.29 | 1405.7 | 2991.81 | 830.68 | 57.97 | 2.71 | 0.17 | 0.02 | 784883.4503 | - |
| | Shortest distance | 8 | 0 | 0 | 0 | 0 | 1508 | 2900 | 787 | 64 | | | | 780132.4535 | - |
| pla33810 | Average distance | 8.22 | 0 | 0 | 0 | 13.16 | 466.54 | 6031.79 | 2798.54 | 108.35 | 0.5 | | | 86349842.36 | - |
| | Shortest distance | 8 | 0 | 0 | 0 | 12 | 489 | 5981 | 2827 | 96 | | | | 85404148.53 | - |

Table A2 Average and Best Distance Ratios of PM & SA against OD and against each other

| City Group | | burma14 | ulysses16 | ulysses22 | bays29 | berlin52 | eil76 | eil101 | gr120 | ch150 | gr202 | tsp225 | a280 | pa561 | gr666 | pr1002 | pr2392 | rl5934 | pla7397 | usa13509 | d18512 | pla33810 |
|---|---|---|---|---|---|---|---|---|---|---|---|---|---|---|---|---|---|---|---|---|---|---|
| **Distance Ratio of Average simulation results** | | | | | | | | | | | | | | | | | | | | | | |
| PM/OD | i5 | 1.05 | 1.05 | 1.05 | 1.06 | 1.15 | 1.13 | 1.13 | 1.11 | 1.15 | 1.04 | 1.18 | 1.23 | 0.92 | 0.94 | 1.22 | 1.26 | 1.35 | 1.29 | - | - | - |
| | i7 | 1.05 | 1.04 | 1.05 | 1.07 | 1.16 | 1.12 | 1.13 | 1.10 | 1.15 | 1.04 | 1.17 | 1.23 | 0.92 | 0.94 | 1.21 | 1.26 | 1.35 | 1.29 | - | - | - |
| | i9 | 1.04 | 1.05 | 1.05 | 1.06 | 1.16 | 1.13 | 1.14 | 1.11 | 1.16 | 1.04 | 1.19 | 1.24 | 0.92 | 0.94 | 1.21 | 1.25 | 1.33 | 1.28 | - | - | - |
| SA/OD | i5 | 1.00 | 1.00 | 1.00 | 0.98 | 1.05 | 1.08 | 1.12 | 1.11 | 1.27 | 1.07 | 1.41 | 1.61 | 1.47 | 1.68 | 3.00 | 6.56 | 17.23 | 19.42 | - | - | - |
| | i7 | 1.00 | 1.00 | 1.00 | 0.98 | 1.05 | 1.08 | 1.12 | 1.12 | 1.27 | 1.08 | 1.40 | 1.62 | 1.47 | 1.68 | 3.01 | 6.55 | 17.21 | 19.39 | - | - | - |
| | i9 | 1.00 | 1.00 | 1.00 | 0.98 | 1.05 | 1.08 | 1.12 | 1.12 | 1.28 | 1.07 | 1.41 | 1.61 | 1.47 | 1.68 | 3.00 | 6.56 | 17.24 | 19.41 | - | - | - |
| **Distance Ratio of Best simulation results** | | | | | | | | | | | | | | | | | | | | | | |
| PM/OD | i5 | 1.00 | 1.01 | 1.00 | 0.99 | 1.04 | 1.07 | 1.07 | 1.05 | 1.07 | 0.94 | 1.12 | 1.18 | 0.90 | 0.90 | 1.17 | 1.23 | 1.32 | 1.27 | - | - | - |
| | i7 | 1.00 | 1.00 | 1.00 | 0.98 | 1.05 | 1.06 | 1.07 | 1.05 | 1.09 | 0.95 | 1.12 | 1.16 | 0.88 | 0.90 | 1.17 | 1.23 | 1.32 | 1.27 | - | - | - |
| | i9 | 1.00 | 1.00 | 1.00 | 0.98 | 1.08 | 1.06 | 1.08 | 1.03 | 1.07 | 0.97 | 1.13 | 1.16 | 0.89 | 0.89 | 1.17 | 1.21 | 1.30 | 1.25 | - | - | - |
| SA/OD | i5 | 1.00 | 1.00 | 1.00 | 0.98 | 1.00 | 1.03 | 1.05 | 1.04 | 1.19 | 1.01 | 1.31 | 1.47 | 1.37 | 1.60 | 2.88 | 6.37 | 16.87 | 19.04 | - | - | - |
| | i7 | 1.00 | 1.00 | 1.00 | 0.98 | 1.00 | 1.03 | 1.06 | 1.06 | 1.18 | 1.01 | 1.28 | 1.45 | 1.36 | 1.59 | 2.87 | 6.35 | 16.91 | 18.92 | - | - | - |
| | i9 | 1.00 | 1.00 | 1.00 | 0.98 | 1.00 | 1.03 | 1.05 | 1.06 | 1.18 | 1.00 | 1.30 | 1.45 | 1.37 | 1.58 | 2.86 | 6.39 | 16.94 | 19.06 | - | - | - |

Table A3 Average and Best Distance Execution time of PM (sequential), PM (parallel processing), and SA implementations

| City | Group | burma14 | ulysses16 | ulysses22 | bays29 | berlin52 | eil76 | eil101 | gr120 | ch150 | gr202 | tsp225 | a280 | pa561 | gr666 | pr1002 | pr2392 | rl5934 | pla7397 | usa13509 | d18512 | pla33810 |
|---|---|---|---|---|---|---|---|---|---|---|---|---|---|---|---|---|---|---|---|---|---|---|
| **Execution time of Average simulation results [seconds]** | | | | | | | | | | | | | | | | | | | | | | |
| SA | i5 | 0.63 | 0.32 | 0.32 | 0.20 | 0.26 | 0.32 | 0.34 | 0.32 | 0.30 | 0.55 | 0.42 | 0.53 | 1.63 | 1.24 | 2.51 | 3.94 | 13.29 | 17.68 | 56.41 | 103.81 | 134.61 |
| | i7 | 0.27 | 0.33 | 0.32 | 0.20 | 0.24 | 0.31 | 0.34 | 0.31 | 0.29 | 0.53 | 0.44 | 0.55 | 1.06 | 1.32 | 1.79 | 4.24 | 13.41 | 18.77 | 47.10 | 66.13 | 137.49 |
| | i9 | 0.63 | 0.32 | 0.32 | 0.20 | 0.26 | 0.32 | 0.34 | 0.32 | 0.30 | 0.55 | 0.42 | 0.53 | 1.63 | 1.24 | 2.51 | 3.94 | 13.29 | 17.68 | 56.41 | 103.81 | 134.61 |
| SPM | i5 | 0.08 | 0.10 | 0.07 | 0.05 | 0.07 | 0.09 | 0.12 | 0.11 | 0.13 | 0.19 | 0.18 | 0.21 | 0.38 | 0.45 | 0.66 | 1.53 | 4.04 | 5.19 | 9.82 | 13.18 | 23.65 |
| | i7 | 0.07 | 0.08 | 0.09 | 0.05 | 0.07 | 0.10 | 0.12 | 0.12 | 0.13 | 0.20 | 0.21 | 0.24 | 0.40 | 0.55 | 0.85 | 1.87 | 4.62 | 5.50 | 10.67 | 14.37 | 25.70 |
| | i9 | 0.08 | 0.10 | 0.07 | 0.05 | 0.07 | 0.09 | 0.12 | 0.11 | 0.13 | 0.19 | 0.18 | 0.21 | 0.38 | 0.45 | 0.66 | 1.53 | 4.04 | 5.19 | 9.82 | 13.18 | 23.65 |
| PPM | i5 | 0.10 | 0.12 | 0.13 | 0.08 | 0.13 | 0.15 | 0.17 | 0.17 | 0.17 | 0.25 | 0.57 | 0.42 | 0.33 | 0.43 | 0.52 | 1.19 | 4.43 | 4.02 | 6.11 | 9.43 | 14.46 |
| | i7 | 0.09 | 0.13 | 0.12 | 0.07 | 0.12 | 0.13 | 0.15 | 0.15 | 0.15 | 0.22 | 0.23 | 0.26 | 0.34 | 0.48 | 0.60 | 0.96 | 2.04 | 2.03 | 4.08 | 4.99 | 9.06 |
| | i9 | 0.10 | 0.12 | 0.13 | 0.08 | 0.13 | 0.15 | 0.17 | 0.17 | 0.17 | 0.25 | 0.57 | 0.42 | 0.33 | 0.43 | 0.52 | 1.19 | 4.43 | 4.02 | 6.11 | 9.43 | 14.46 |
| **Execution time of Best simulation results [seconds]** | | | | | | | | | | | | | | | | | | | | | | |
| SA | i5 | 0.88 | 0.30 | 0.33 | 0.17 | 0.38 | 0.34 | 0.38 | 0.33 | 0.31 | 0.55 | 0.44 | 0.63 | 1.09 | 1.25 | 4.37 | 3.79 | 11.15 | 49.16 | 38.09 | 57.04 | 122.74 |
| | i7 | 0.24 | 0.28 | 0.30 | 0.22 | 0.26 | 0.37 | 0.37 | 0.37 | 0.34 | 0.55 | 0.49 | 0.71 | 1.15 | 1.38 | 1.80 | 4.15 | 13.10 | 18.37 | 46.47 | 66.39 | 136.79 |
| | i9 | 0.88 | 0.30 | 0.33 | 0.17 | 0.38 | 0.34 | 0.38 | 0.33 | 0.31 | 0.55 | 0.44 | 0.63 | 1.09 | 1.25 | 4.37 | 3.79 | 11.15 | 49.16 | 38.09 | 57.04 | 122.74 |
| SPM | i5 | 0.17 | 0.03 | 0.05 | 0.06 | 0.05 | 0.08 | 0.22 | 0.08 | 0.09 | 0.17 | 0.13 | 0.17 | 0.34 | 0.44 | 0.65 | 1.46 | 3.78 | 5.32 | 9.90 | 13.20 | 23.16 |
| | i7 | 0.17 | 0.11 | 0.02 | 0.05 | 0.06 | 0.15 | 0.09 | 0.14 | 0.12 | 0.18 | 0.22 | 0.22 | 0.41 | 0.57 | 0.80 | 1.86 | 4.49 | 5.45 | 10.56 | 14.34 | 25.65 |
| | i9 | 0.17 | 0.03 | 0.05 | 0.06 | 0.05 | 0.08 | 0.22 | 0.08 | 0.09 | 0.17 | 0.13 | 0.17 | 0.34 | 0.44 | 0.65 | 1.46 | 3.78 | 5.32 | 9.90 | 13.20 | 23.16 |
| PPM | i5 | 0.22 | 0.22 | 0.06 | 0.09 | 0.09 | 0.14 | 0.20 | 0.23 | 0.16 | 0.31 | 0.50 | 0.64 | 0.34 | 0.41 | 0.53 | 1.08 | 2.68 | 2.72 | 6.15 | 6.32 | 14.54 |
| | i7 | 0.22 | 0.04 | 0.06 | 0.05 | 0.11 | 0.18 | 0.12 | 0.15 | 0.14 | 0.36 | 0.22 | 0.26 | 0.34 | 0.50 | 0.54 | 0.86 | 2.09 | 2.05 | 4.12 | 5.03 | 9.22 |
| | i9 | 0.22 | 0.22 | 0.06 | 0.09 | 0.09 | 0.14 | 0.20 | 0.23 | 0.16 | 0.31 | 0.50 | 0.64 | 0.34 | 0.41 | 0.53 | 1.08 | 2.68 | 2.72 | 6.15 | 6.32 | 14.54 |

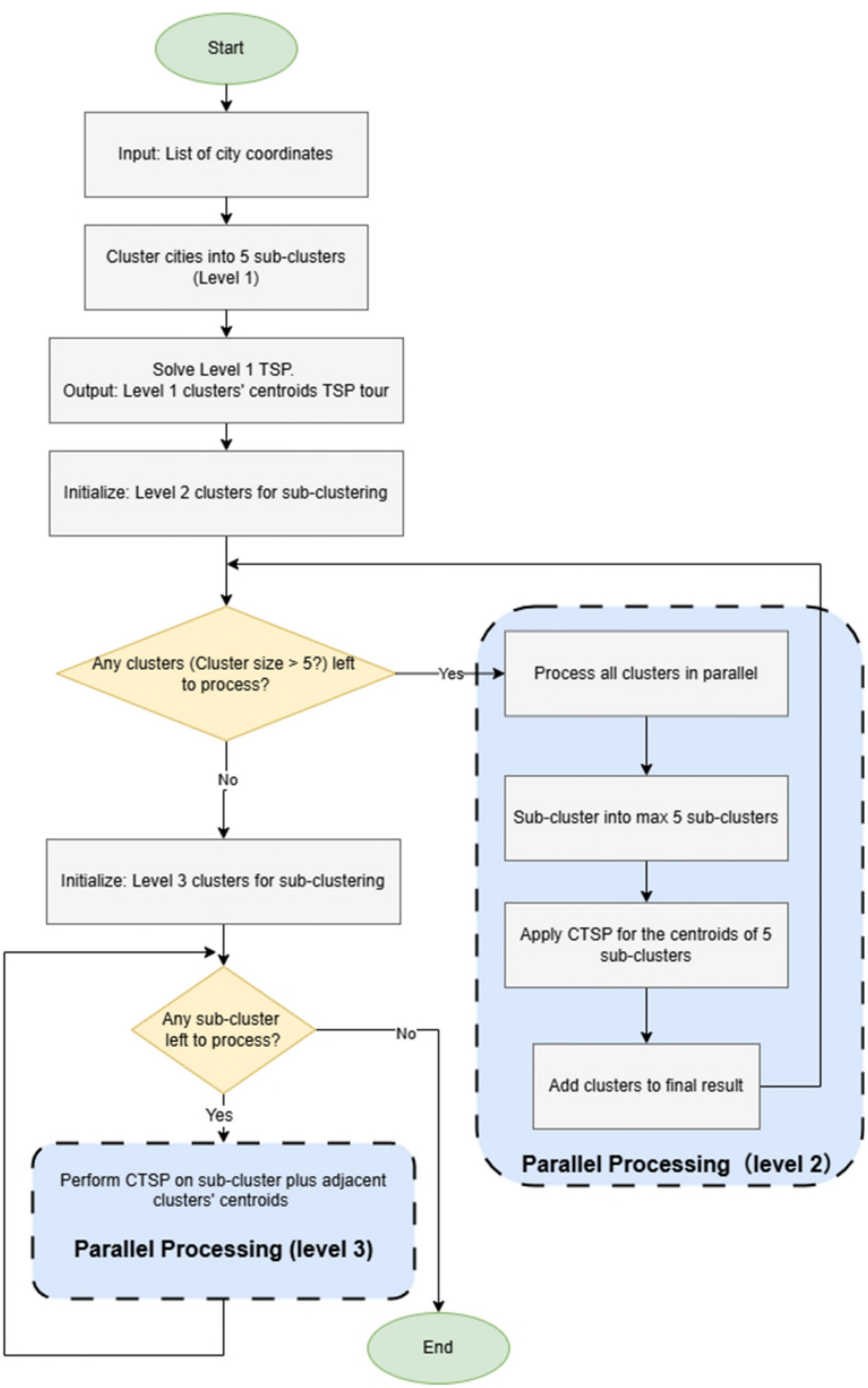

Figure A1 Flowchart of the overall/main Proposed Method (Algorithm A1)

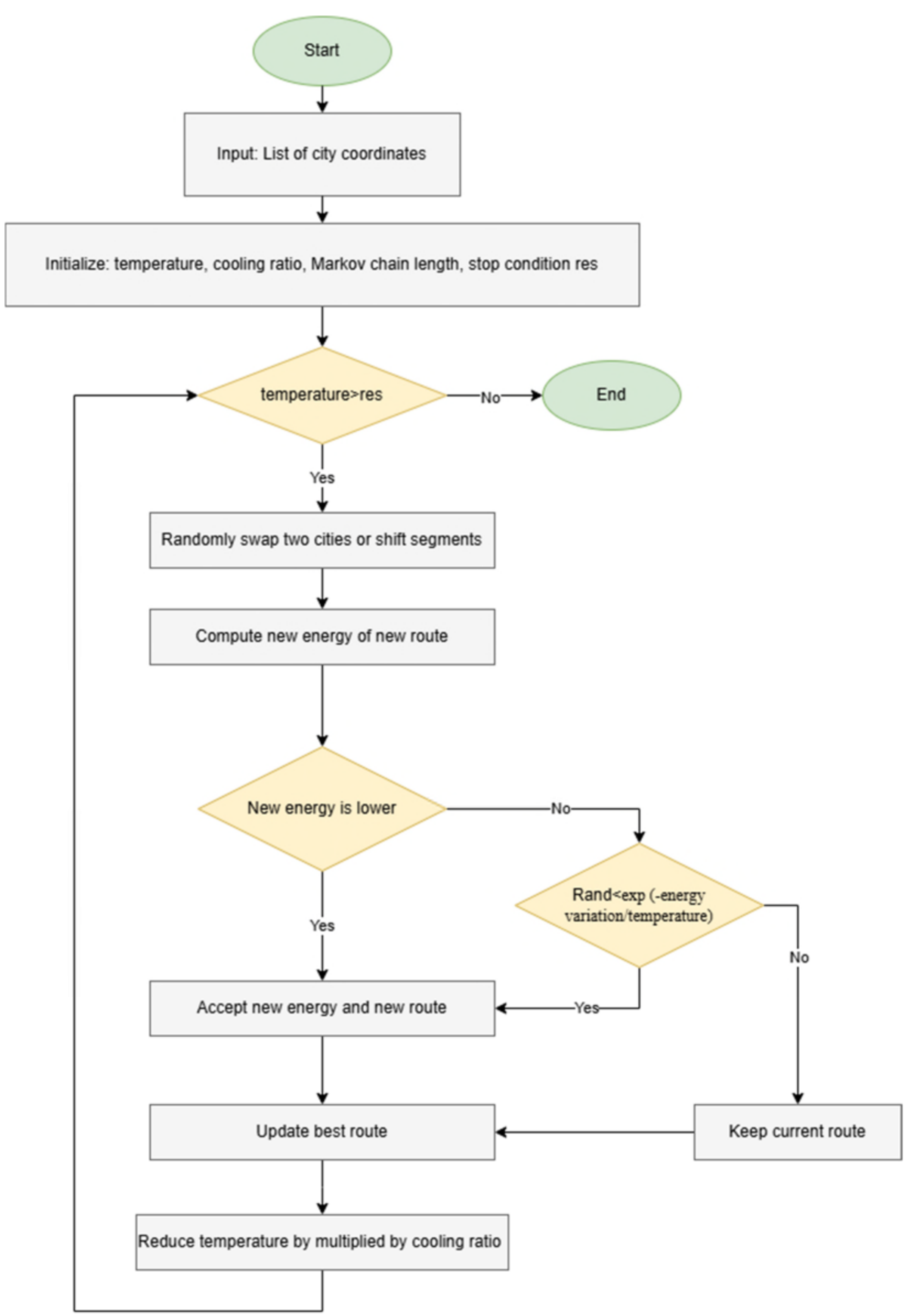

Figure A2 Flowchart of the Simulated Annealing TSP algorithm, used in Level 1.

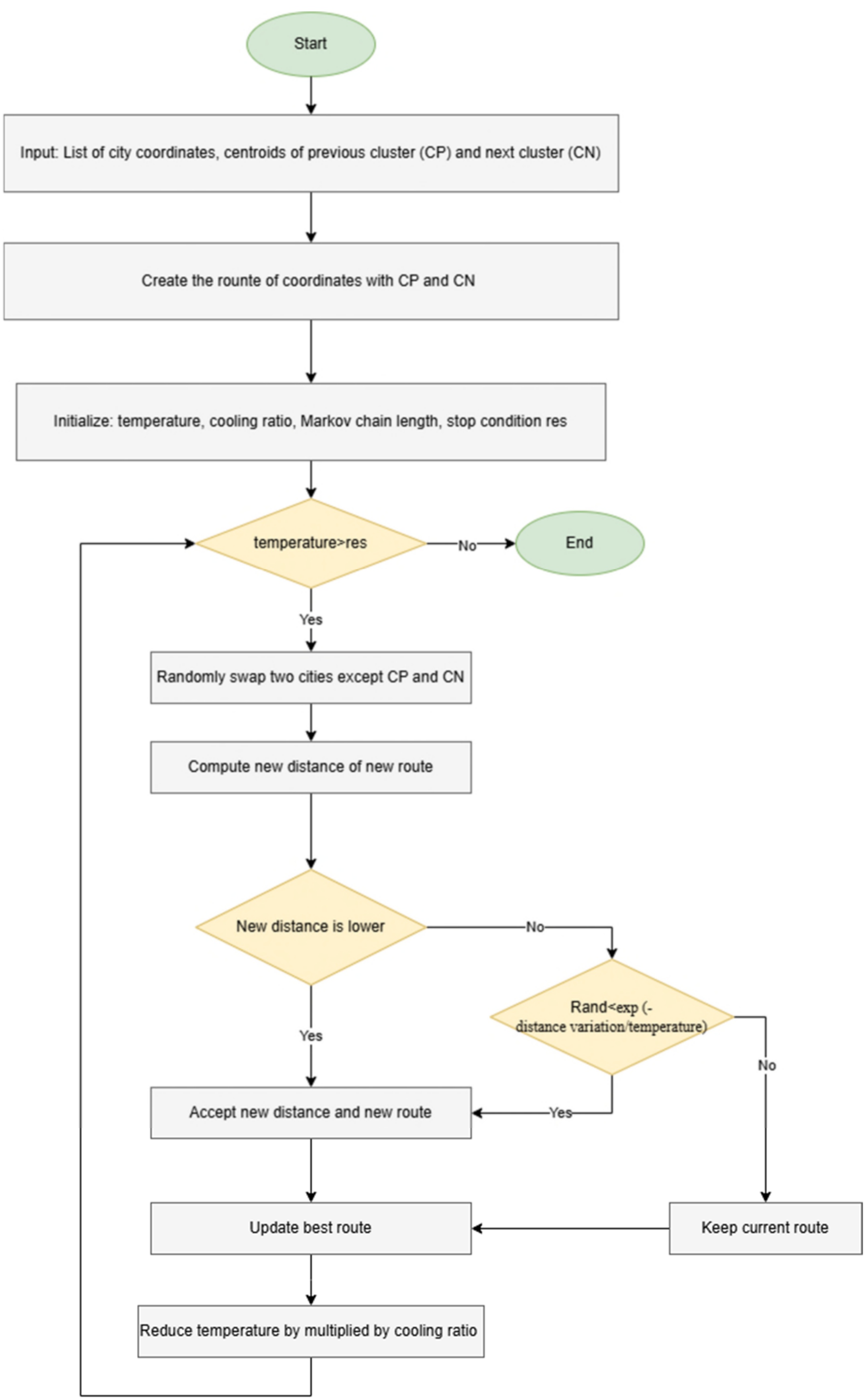

Figure A3 Flowchart of the Simulated Annealing Constrained TSP (CTSP) algorithm, used in Level 2 onwards.